\documentclass[preprint,12pt]{elsarticle}

\usepackage[T1]{fontenc}

\usepackage{graphicx}
\graphicspath{./}

\usepackage{amssymb}

\usepackage{amsmath}

\usepackage[scientific-notation=false]{siunitx}

\usepackage{amsthm}
\newtheorem{definition}{Definition}

\counterwithin{figure}{section}
\counterwithin{table}{section}
\counterwithin{definition}{section}
\counterwithin{equation}{section}

\usepackage{setspace}

\usepackage{rotating}
\usepackage{tabularx}

\usepackage{xcolor}
\definecolor{ACMblue}{RGB}{1,130,172}
\definecolor{ACMyellow}{RGB}{255,214,0}
\definecolor{ACMorange}{RGB}{252,146,0}
\definecolor{ACMred}{RGB}{253,27,20}
\definecolor{ACMlightblue}{RGB}{131,206,226}
\definecolor{ACMgreen}{RGB}{166,188,9}
\definecolor{ACMpurple}{RGB}{101,1,107}
\definecolor{ACMdarkblue}{RGB}{9,53,122}
\definecolor{GPTtealblue}{RGB}{0, 128, 128}
\definecolor{GPTburntorange}{RGB}{204, 85, 0}
\definecolor{forestgreen}{RGB}{34, 139, 34}
\definecolor{GPTdeeplavender}{RGB}{128, 0, 128}
\definecolor{GPTgoldenrodyellow}{RGB}{218, 165, 32}
\newcommand{\colorACA}{\textcolor{ACMred}} 
\newcommand{\colorSCA}{\textcolor{ACMgreen}}
\newcommand{\colorMSCA}{\textcolor{ACMorange}}
\newcommand{\colorENCA}{\textcolor{ACMpurple}} 
\newcommand{\colorNUCA}{\textcolor{ACMblue}}

\usepackage{booktabs}

\usepackage{array}
\newcommand{\PreserveBackslash}[1]{\let\temp=\\#1\let\\=\temp}
\newcolumntype{C}[1]{>{\PreserveBackslash\centering}p{#1}}
\newcolumntype{R}[1]{>{\PreserveBackslash\raggedleft}p{#1}}
\newcolumntype{L}[1]{>{\PreserveBackslash\raggedright}p{#1}}

\usepackage{multicol}

\journal{CNSNS} 

\begin{document}

\begin{frontmatter}



\title{A Comprehensive Taxonomy of Cellular Automata}


\author[ugent]{Michiel Rollier}
\ead{michiel.rollier@ugent.be}

\author[usp]{Kallil M.C.~Zielinski}

\author[ugent]{Aisling J.~Daly}

\author[usp]{Odemir M.~Bruno}

\author[ugent]{Jan M.~Baetens}

\affiliation[ugent]{
  organization={BionamiX, Fac.~of Bioscience Engineering, Ghent University},
  addressline={Coupure Links, 653},
  city={Ghent},
  postcode={9000},
  country={Belgium}
}

\affiliation[usp]{
  organization={S\~ao Carlos Institute of Physics, University of S\~ao Paulo},
  addressline={Av.~Trab.~S\~ao Carlense, 400},
  city={S\~ao Carlos},
  postcode={13566-590},
  country={Brazil}
}

\begin{abstract}
Cellular automata (CAs) are fully-discrete dynamical models that have received much attention due to the fact that their relatively simple setup can nonetheless express highly complex phenomena.
Despite the model's theoretical maturity and abundant computational power, the current lack of a complete survey on the `taxonomy' of various families of CAs impedes efficient and interdisciplinary progress.
This review paper mitigates that deficiency; it provides a methodical overview of five important CA `families': asynchronous, stochastic, multi-state, extended-neighbourhood, and non-uniform CAs.
These five CA families are subsequently presented from four angles. First, a rigorous  mathematical definition is given. Second, we map prominent variations within each CA family, as such highlighting mathematical equivalences with types from other families. Third, we discuss the genotype and phenotype of these CA types by means of mathematical tools, indicating when established tools break down. Fourth, we conclude each section with a brief overview of applications related to information theory and mathematical modelling.

\end{abstract}




\end{frontmatter}


\section{Introduction}
\label{sec:introduction}

\subsection{Definition}
\label{subsec:definition}

A cellular automaton (CA) is a discrete dynamical system, first defined by John von Neumann while attempting to abstractly mimic and explain biological self-replication \cite{neumann1966theory}. He describes an $n$-dimensional ($n$-D) infinite regular lattice whose cells can take a discrete number of states. The state of every cell is evaluated at regular time steps, obeying a rule that only depends on its own state and the states of cells in its proximity. We represent such an object as a sextuple $\mathcal{C}$ as follows.

\begin{definition}[Cellular automaton]
\label{def:CA_von-neumann}
A cellular automaton (CA) is a sextuple $\mathcal{C} = \langle \mathcal{T}, S, s, s_0, \mathcal{N}, \phi \rangle$, where
\begin{enumerate}
    \item $\mathcal{T}$ is a countably infinite regular tessellation of an $n$-dimensional Euclidean space $\mathbb{R}^n$ consisting of cells $c_i$, $i \in \mathbb{Z}$;
    \item $S$ is a finite set of $k$ states, often $S \subset \mathbb{Z}$
    \item The output function $s : \mathcal{T} \times \mathbb{Z} \rightarrow S$ yields the state value of cell $c_i$ at the $t$-th discrete time step, i.e.~$s(c_i, t)$;
    \item The function $s_0 : \mathcal{T} \rightarrow S$ assigns to every cell $c_i$ an initial state, i.e.~$s(c_i, 0) = s_0(c_i)$;
    \item The neighbourhood function $\mathcal{N} : \mathcal{T} \rightarrow \mathcal{P}(\mathcal{T})$ maps every cell $c_i \in \mathcal{T}$ to an element in the power set $\mathcal{P}(\mathcal{T})$, i.e.~a finite subset of $\mathcal{T}$, such that $$\mathcal{N}(c_i) = \{c_{i,1}, \ldots, c_{i,{|\mathcal{N}|}}\}\equiv \{ c_{i,j}\}_{j=1}^{|\mathcal{N}|},$$ where $c_{i,j}$ is the $j$-th neighbour of cell $c_i$.
    \item The function $\phi : S^{|\mathcal{N}|} \rightarrow S$ determines the next state of cell $c_i$, i.e.~$$s(c_i, t+1) = \phi(\tilde{s}(\mathcal{N}(c_i),t)),$$ where $\tilde{s}(\mathcal{N}(c_i),t)$ is the tuple $\left(s(c_{i,j},t)\right)^{|\mathcal{N}|}_{j=1}$. This function $\phi$ is called the `local update rule'.
\end{enumerate}
By acting on all cells, $\phi$ induces a `global update rule' $\Phi : S^{\mathbb{Z}^n} \rightarrow S^{\mathbb{Z}^n}$ on the space of all possible configurations.
\end{definition}

Despite their simple nature, CAs are capable of generating spectacular complexity; some have even been proven to be computationally universal \cite{Wolfram1984}. The potential of this ``new kind of science'' \cite{Wolfram2002} has lead to an increasingly wide interest from scholars over the entire scientific landscape \cite{sarkar2000history}, roughly classified into three axes of study \cite{fates2014}: pure mathematics, computer science, and mathematical modelling. While the maths of CAs is concerned with formal CA structures and symbolic dynamics \cite{kari2005theory}, the latter two axes of study are associated with a number of applications, a selection of which we list in Tab.~\ref{tab:list-of-ca-research}. In virtually all of these domains, however, one or more of the defining items of the CA `sensu stricto' (cf.~Def.~\ref{def:CA_von-neumann}) are generalised. This results in abandoning spatiotemporal uniformity, locality, and/or synchronicity \cite{bhattacharjee2020survey}, and allows for what we may call CAs `sensu lato'.

\begin{sidewaystable}
    \centering
    \caption{An overview of various research domains in which CAs are used, demonstrated by an illustrative (though not exhaustive) selection of references to (topical) surveys, influential classic papers, and more recent work. Citations are colour-coded according to the CA family the application principally makes use of (\colorACA{ACA}, \colorSCA{SCA}, \colorMSCA{MSCA}, \colorENCA{ENCA}, \colorNUCA{$\nu$CA}, cf.~Tab.~\ref{tab:nECAs}). Citations coloured black either use ECAs, or are not easily identified with a single CA family.}
    \begin{tabular}{L{4.2cm}L{4.2cm}L{4.2cm}L{4.2cm}}
        \textbf{Research domain} & \textbf{Surveys} & \textbf{Classic papers} & \textbf{Recent work} \\ \toprule
        \multicolumn{4}{l}{\textsc{Computer science}} \\
        \ Artificial life & \cite{ilachinski2001cellular} & \colorMSCA{\cite{langton1984self,langton1986al,stauffer1998relationship}} & \colorENCA{\cite{pena2021life}} \\
        \ Computer architecture & \cite{monica2016cellular} & \colorENCA{\cite{biafore1994cellular,khan1997vlsi}} & \cite{liu2023memristor}, \colorSCA{\cite{yamamoto2021statica}}, \colorENCA{\cite{cagigas2022efficient}} \\
        \ Computational tasks & \cite{Mitchell2000,mitchell2005computation} & \colorACA{\cite{szabo1998prisoner}}, \colorNUCA{\cite{sipper1996}} & \colorACA{\cite{ruivo2019perfect}}, \colorMSCA{\cite{mordvintsev2020growing,palm2022variational}}\\
        \ Cryptography & \cite{poornima2017prng} & \cite{wolfram1986random,abdo2013crypto}, \colorENCA{\cite{lafe1997compression}}, \colorNUCA{\cite{sen2002crypto}} & \cite{mondal2019secure}, \colorENCA{\cite{wang2018image,su2019reversible,coronabermudez2022cryptographic}}\\
        \ Data compression & \cite{rosin2014cellular} & \colorMSCA{\cite{culik1993image}}, \colorENCA{\cite{lafe1997compression}}, \colorNUCA{\cite{nara1999novel,wada2002completely}} & \cite{chai2020efficient,zhang2020novel,ahmed2020using}, \colorENCA{\cite{milani2017fast}} \\
        \ Pattern recognition & \cite{maji2003theory} & \cite{mylopoulos1972application}, \colorMSCA{\cite{maji2005fuzzy}} & \colorENCA{\cite{florindo2021texture}} \\ \midrule 
        \multicolumn{4}{l}{\textsc{Mathematical modelling}} \\
        \ Biology & \cite{ermentrout1993bio,alber2003bio,bonchev2010biomolecular,xiao2011bioinformatics} & \colorACA{\cite{kier1999biochemical}}, \colorSCA{\cite{inghe1989genet}}, \colorNUCA{\cite{sieburg1990hiv,pandey1991cellular}} & \colorMSCA{\cite{mordvintsev2020growing,khaleghi2021neuronal}}, \colorENCA{\cite{manukyan2017living}} \\
        \ Chemistry & \cite{raabe2002material,menshutina2020chemistry} & \colorACA{\cite{seybold1997kinetics,kier1999biochemical,kier2000cellular}}, \colorSCA{\cite{chopard1991diffusion,rappaz1993probabilistic}}, \colorMSCA{\cite{zanette1992multistate}}, \colorENCA{\cite{weimar1992third}} & \colorSCA{\cite{yanez2020evaluation,gong2022nucleation}}, \colorMSCA{\cite{tsompanas2021belousov}} \\
        \ Ecology & \cite{hogeweg1988eco,phipps1992local,balzter1998vegetation}, \colorSCA{\cite{durrett1994stochastic}} & \colorMSCA{\cite{hassell1991insect}}, \colorENCA{\cite{hendry1995memory}}, \colorNUCA{\cite{rand1995eco}} & \colorSCA{\cite{buschmann2023cost}}, \colorMSCA{\cite{mi2023detecting}} \\
        \ Epidemiology & \cite{hadeler2016epidemiology,delrey2015malware} & \colorSCA{\cite{fuentes1999epi,mikler2005modeling,medeiros2011dengue}}, \colorMSCA{\cite{rhodes1997epidemic}} & \colorSCA{\cite{ghosh2020covid,mondal2020epi,jithesh2021covid,schimit2021brazil,lu2023spatial}}, \colorMSCA{\cite{cavalcante2021covid}}, \colorENCA{\cite{nava2020seir}}, \colorNUCA{\cite{dai2021epidemics,medrek2021corona,eosina2022covid}}\\
        \ Geology & \cite{plotnick2016lattice,jimenez2013seismicity} & \colorMSCA{\cite{henderson1994fracture,akishin1998simulation}}, \colorENCA{\cite{coulthard2006river}} & \colorMSCA{\cite{kerin2022mountain}}, \colorENCA{\cite{machado2015lahars}}, \colorNUCA{\cite{caracciolo2014climate}} \\
        \ Medical science & \colorNUCA{\cite{Ribba2004}} & \colorSCA{\cite{celada1992immune}}, \colorMSCA{\cite{neumann1989immune,pandey1991cellular}}, \colorENCA{\cite{kansal2000simulated}}, \colorNUCA{\cite{alarcon2003tumour}} & \colorMSCA{\cite{nejad2014fuzzy,luna-benoso2022melanoma}}, \colorENCA{\cite{hadavi2014lung,fan2019lossless}} \\
        \ Forest fires & \cite{sullivan2009wildland,papadopoulos2011wildfire} & \colorENCA{\cite{clarke1994wildfire,hargrove2000fire}} & \colorSCA{\cite{mutthulakshmi2020simulating,li2022lstm}}, \cite{zheng2017forest,freire2019using} \\
        \ Physics & \cite{chopard1998physical,ilachinski2001cellular}, \colorENCA{\cite{boghosian1999lattice,ilachinski2001cellular}} & \cite{minsky1982vacuum,vichniac1984}, \colorSCA{\cite{domany1984equivalence}}, \colorMSCA{\cite{toffoli1984alternative}}, \colorENCA{\cite{wolfram1986fluids,lejeune1999nbody}} & \cite{buca2021rule}, \colorMSCA{\cite{iadecola2020nonergodic,kerin2022mountain}} \\
        \ Sociology & \cite{hegselmann1996social} & \colorSCA{\cite{lewenstein1992statistical,nowak1996social}} & \colorACA{\cite{silva2020information}}, \colorSCA{\cite{waldorp2020mean}} \\
        \ Traffic & \cite{wolf1999traffic,maerivoet2005road}, \colorNUCA{\cite{gao2020urban}} & \colorSCA{\cite{schadschneider2002traffic,nagel1992freeway}} & \colorENCA{\cite{maecki2017graph}}, \colorNUCA{\cite{padovani2018modeling}} \\
        \ Urban planning & \cite{Sante2010,aburas2016simulation,Chakraborty2022} & \colorENCA{\cite{white1993urban}}, \colorNUCA{\cite{clarke1997self,clarke1998loose}} & \colorENCA{\cite{liao2016incorporation}}, \colorMSCA{\cite{xu2022forecasting}}, \colorNUCA{\cite{jamali2019flood}} \\ \bottomrule
    \end{tabular}
    \label{tab:list-of-ca-research}
\end{sidewaystable}

The goal of this review paper is to map the taxonomy of the most influential CA variations that go beyond the classical definition in Def.~\ref{def:CA_von-neumann}, additionally indicating the implications of the particular extension on the mathematical structure and behaviour, and listing a number of prominent applications. Rather than an in-depth discussion of the applications, or a lengthy collection of scrutinous formal proofs, our ambition is to establish an overview of the CA landscape \textit{starting} from the previous definition of a classical CA.

While Bhattacharjee et al.~\cite{bhattacharjee2020survey} established a similar survey, the merit of our work lies in its highly systematic approach. Our hope is that this will help the reader to observe the large collection of CA extensions not as a melting pot of trial and error, but as a well-structured family tree. Like a taxonomist, we identify various CA families, each containing several types with distinct definition variations. Unlike in biological taxonomy, however, we encounter CA types that belong to multiple families, due to a demonstrated mathematical equivalence.

\subsection{Elementary CAs}
\label{subsec:elementary-cas}
\begin{figure}
    \centering
    \includegraphics[height=.2\linewidth]{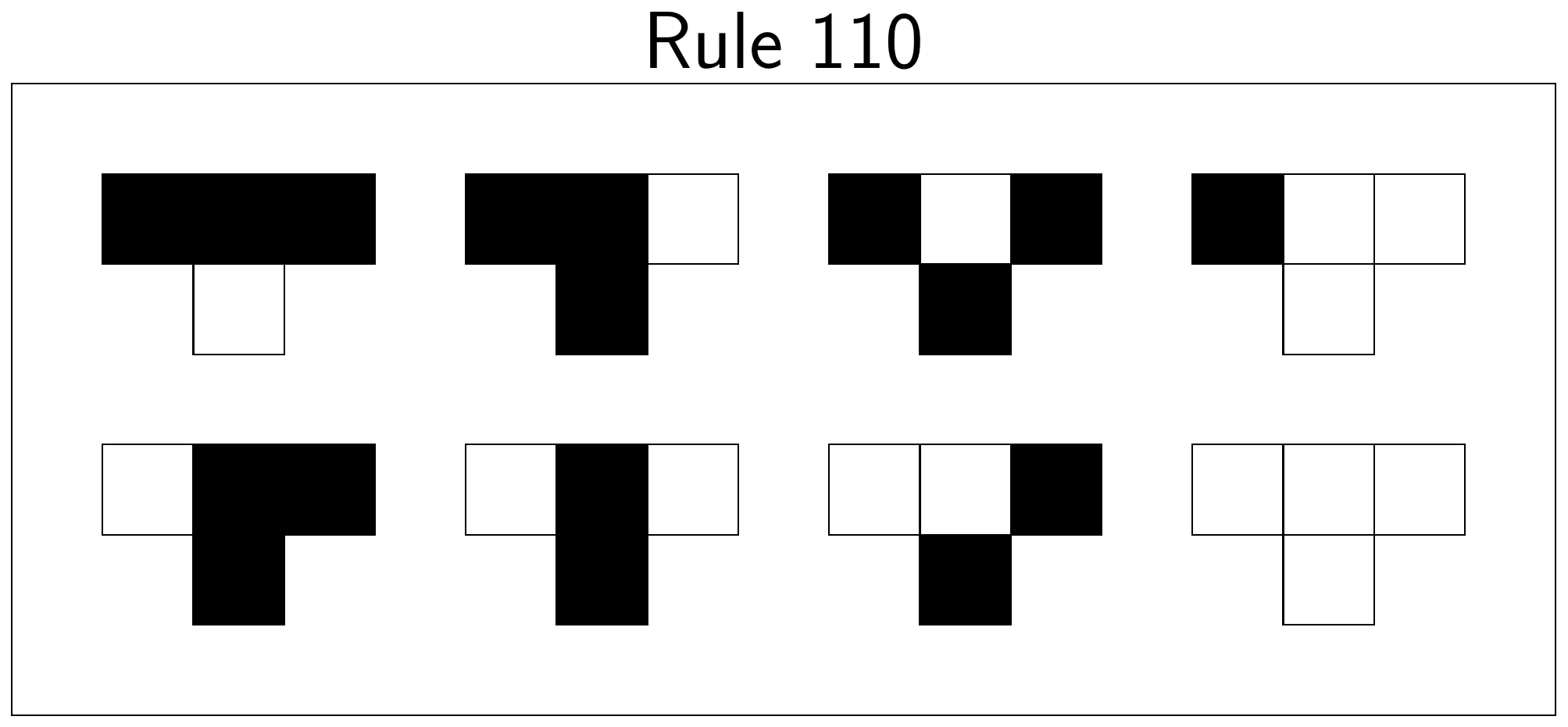}\qquad\qquad
    \includegraphics[height=.2\linewidth]{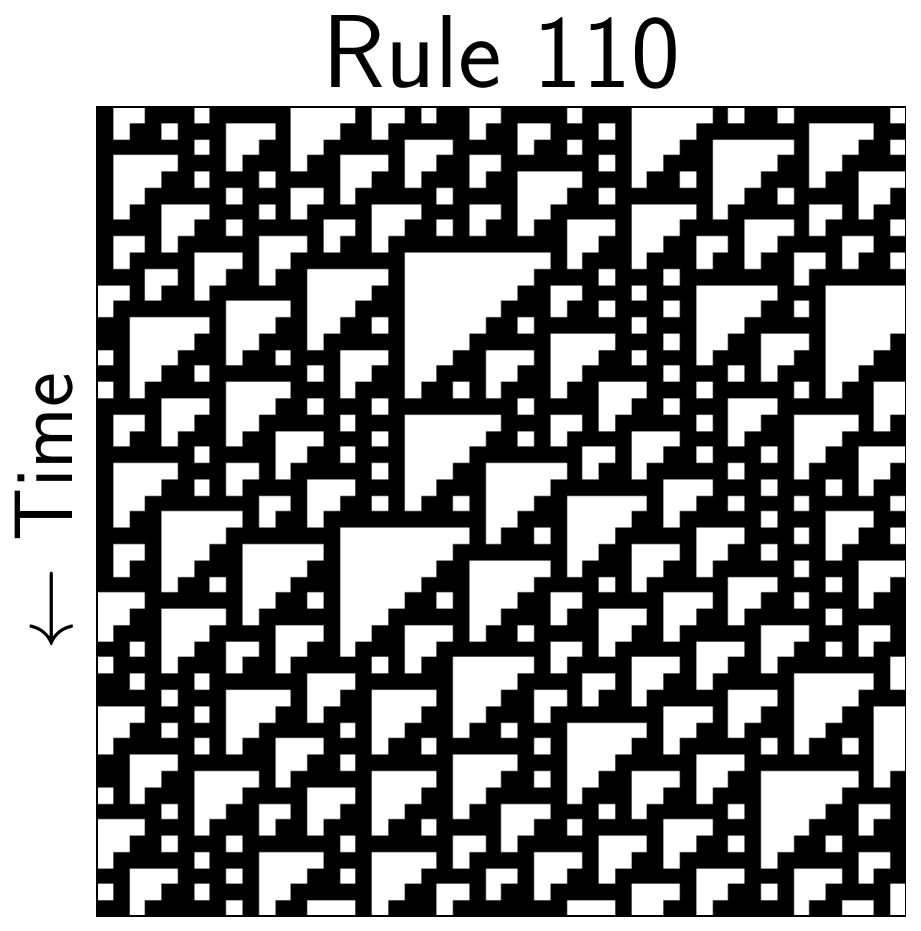}
    \caption{Rule table (left) and a spacetime diagram (right) for elementary rule 110, which can generate complex behaviour \cite{Cook2004}, despite its minimal set-up.}
    \label{fig:rule_110_rule_table}
\end{figure}

One type of CA sensu stricto, the so-called `elementary' CA (ECA), has been studied extensively \cite{Wolfram1984,Wolfram1983,Li1990}, and serves as the CA archetype. This is due to the observation that such CAs are the simplest non-trivial type (leaving 1-neighbour binary CAs aside \cite{bandini2012}). The ECA is defined by further narrowing Def.~\ref{def:CA_von-neumann} as follows:

\begin{definition}[Elementary cellular automaton]
\label{def:ECA}
An elementary cellular automaton (ECA) is a sextuple $\mathcal{C} = \langle \mathcal{T}, S, s, s_0, \mathcal{N}, \phi \rangle$, where
\begin{enumerate}
    \item $\mathcal{T}$ is a countably infinite regular tessellation of the 1-D Euclidean space $\mathbb{R}$ consisting of cells $c_i$, $i \in \mathbb{Z}$;
    \item $S$ is the binary set $\{0,1\}$;
    \item The output function $s : \mathcal{T} \times \mathbb{Z} \rightarrow S$ yields the state value of cell $c_i$ at the $t$-th discrete time step, i.e.~$s(c_i, t)$;
    \item The function $s_0 : \mathcal{T} \rightarrow S$ assigns to every cell $c_i$ an initial state, i.e.~$s(c_i, 0) = s_0(c_i)$;
    \item The neighbourhood function $\mathcal{N} : \mathcal{T} \rightarrow \mathcal{T}^3$ maps every cell $c_i$ to the set $\mathcal{N}(c_i) = \{c_{i-1}, c_i, c_{i+1}\};$
    \item The local update rule $\phi : S^3\rightarrow S$ determines the cell state evolution, i.e.~$s(c_i, t+1) = \phi(s(c_{i-1},t), s(c_i,t), s(c_{i+1},t)).$
\end{enumerate}
\end{definition}
\noindent This setup allows for $2^{2^3} = 256$ local update rules $\phi$ commonly referred to as elementary or `Wolfram' rules, and denoted by an integer $W(\phi)$ from 0 to 255 calculated on the basis of the rule table \cite{Wolfram2002}. We observe two symmetries: the CA's overall dynamics remain the same when taking the mirror image (`reflection'), and upon switching states (`complement'). This implies that we are left with only 88 independent rules (see e.g.~\cite{Li1990}). Some of these very simple rules are still capable of generating complexity \cite{Cook2004}, e.g.~rule 110 (Fig.~\ref{fig:rule_110_rule_table}).

Practically all CA-related research outside pure mathematics considers a finite tessellation $\mathcal{T}^* \subset \mathcal{T}$ consisting of $N$ cells $\{c_i\ |\ 0 < i  \leq N\}$, in order to allow for `in silico' simulations. It is then required to specify the neighbourhood of boundary cells $c_1$ and $c_N$, resulting in a slight alteration of Def.~\ref{def:ECA}:
\begin{definition}[Finite ECA with boundary conditions]
\label{def:ECA_BCs}
A finite ECA with $N$ cells and boundary conditions (BCs) is a sextuple $\mathcal{C} = \langle \mathcal{T}^*, S, s, s_0, \mathcal{N}, \phi \rangle$, as defined in Def.~\ref{def:ECA}, with changes to properties 1 and 5:
\begin{enumerate}
    \item[1. ] $\mathcal{T}^*$ is a finite regular tessellation of $\mathbb{R}$ consisting of $N$ cells $\{c_1, \ldots, c_N\}$;
    \item[5. ] The neighbourhood function $\mathcal{N}: \mathcal{T}^* \rightarrow (\mathcal{T}^*)^3 \cup \{c_0, c_{N+1}\}$ maps every internal cell $c_i$ to the set $\mathcal{N}(c_i) = \{c_{i-1}, c_i, c_{i+1}\}$.
\end{enumerate}
\end{definition}
\noindent We rename $\mathcal{T}^* \rightarrow \mathcal{T}$ for notational convenience and assume a finite number of cells in this survey, unless stated otherwise. The BCs specify the state of `fictitious' cells $c_0$ and $c_{N+1}$, most commonly identifying $s(c_0)=s(c_N)$ and $s(c_{N+1}) = s(c_1)$ (periodic boundary conditions). We provide an overview of possible BCs in Tab.~\ref{tab:boundary-conditions}. The choice for particular BCs can significantly affect the dynamics of the system, in particular the propagation of information and the emergence of patterns \cite{Luvalle2019}.

\begin{sidewaystable}
    \centering
    \caption{Overview of possible 1-D ECA boundary conditions (BCs), based on LuValle \cite{Luvalle2019} and Bhattacharjee et al.~\cite{bhattacharjee2020survey}.}
    \begin{tabular}{lL{5cm}p{10cm}}
        BC type & Definition & Description \\ \toprule
        Null & N/A & The boundary is not defined and only cells that have a meaningful neighbourhood are updated. Information is lost and the CA is finite in time.\\
        Fixed & $s(c_0,t) = s_0$, $s(c_{N+1},t) = s_{N+1}$, $s_0,s_{N+1} \in S$ & The partially fixed state of neighbourhoods of $c_1, c_N$ generally decreases the number of possible local update results. \\
        Periodic & $s(c_0) = s(c_{N})$, $s(c_{N+1}) = s(c_1)$ & Synonymous to  ``cyclic'' BCs. No cell `sees' the boundary. Conceptually and mathematically close to an infinite $\mathcal{T}$. \\
        Adiabatic & $s(c_0) = s(c_1)$, $s(c_{N+1}) = s(c_N)$ & Related to Neumann BCs for numerical solutions of (P)DEs. Decreases the number of possible update values. \\
        Reflexive & $s(c_0) = s(c_2)$, $s(c_{N+1}) = s(c_{N-1})$ & Also related to Neumann BCs for numerical solutions of (P)DEs. Decreases the number of possible update values. \\
        Intermediate & $s(c_0) = s(c_i)$, $s(c_{N+1}) = s(c_j)$ for some $i,j \in \{1, \ldots, N\}$ & Generalisation of previous BCs. For symmetry, typically $i + j = M + 1$. If  $2 < i,j < M - 1$, all possible state configurations are still possible. \\
        Random & $s(c_0, t), s(c_{N+1}, t) \in S$ & Entirely random choice for every time step $t$. \\
        Distributed & $s(c_0,t), s(c_{N+1},t) \sim P(S)$ & Most general case: a random choice is made for a particular probability distribution over $S$.\\ \bottomrule
    \end{tabular}
    \label{tab:boundary-conditions}
\end{sidewaystable}

ECAs have historically gotten attention as random number generators \cite{wolfram1986random}, for inspiring computer processor design \cite{Pries1986}, and they have served as toy models, e.g.~in traffic modelling \cite{rosenblueth2011model}. Occasionally, the effect of BCs on such mathematical models is explicitly investigated, e.g.~in pedestrian flow models \cite{Zhang2014}. It should be clear, however, that the number of practical applications of ECAs is limited, which motivates extensions to non-classical CAs. The five CA families discussed in this paper will be defined as a (often minor) variation on Def.~\ref{def:ECA}, tacitly including the possibility of adding BCs such as in Def.~\ref{def:ECA_BCs}.

\subsection{CA properties as taxonomic tools}
Research into the nature and behaviour of CAs has generated a large amount of jargon denoting particular properties that allow for CA classification. In a recent survey, Vispoel et al.~\cite{Vispoel2022} distinguishes four approaches for characterising CAs: (i) classification based on the rule table, (ii) identification of global properties, (iii) a division into behavioural classes, and (iv) quantification based on local properties. The former two may be called `genotypic' (Tab.~\ref{tab:list-of-math-tools-genotypic}), and the latter `phenotypic' (Tab.~\ref{tab:list-of-math-tools-phenotypic}), because they resp.~take information from the model setup (definition) and the model outcome (simulation).

\begin{sidewaystable}
    \centering
    \caption{Mathematical tools and concepts used for characterising and classifying CAs based on the genotype (model definition). The first reference indicates the original academic work. Selection largely copied from Vispoel et al.~\cite{Vispoel2022}.}
    \begin{tabular}{L{.25\linewidth}p{.6\linewidth}L{.15\linewidth}}
        \textbf{Tool/concept} & \textbf{Brief description} & \textbf{References} \\ \toprule
        \multicolumn{3}{l}{\textsc{Rule table parameters}} \\
        \ Langton's parameter & Fraction $\lambda$ of neighbourhood configurations that are not mapped to the quiescent state. & \cite{Langton1990} \\
        \ Mean-field parameters & Generalisation of Langton's $\lambda$. & \cite{Gutowitz1987}; \cite{Li1990,Gutowitz1990} \\
        \ $Z$-reverse parameter & Measure for number of possible predecessors of a state configuration. & \cite{wuensche1992global}; \cite{Langton1990,Wuensche2009,Wuensche2010} \\
        \ $\mu$-sensitivity & The average number of changes in the local update function's output when changing the state of a single neighbourhood cell. & \cite{DeOliveira2000guidelines}; \cite{binder1993phase} \\
        \ Obstruction parameter & The fraction of neighbourhood pairs for which the CA rule is additive. & \cite{voorhees1997some}; \\
        \ Kolmogorov complexity & Entropy-related measure of rule table complexity. Often estimated by Lempel-Ziv complexity. & \cite{Kolmogorov1963}; \cite{ziv1977universal,blanchard2005chaotic,israeli2006coarse} \\ \midrule
        \multicolumn{3}{l}{\textsc{Global properties}} \\
        \ Attractor structure & Investigates the structure between all possible subsequent configurations. & \cite{hurley1990attractors}; \cite{martin1984algebraic,wuensche1992global,hanson1992attractor} \\
        \ Topological dynamics & Applies a number of concept from dynamical systems theory (e.g.~topological concepts such as equicontinuity, sensitivity, transitivity, and positive expansivity) to the space of all possible CA configurations. & \cite{kurka1997languages}; \cite{martin1984algebraic,schule2012full,dennunzio2013periodic,dennunzio2014multidimensional} \\
        \ Formal language theory & Computational approach to studying CA attractors. Configurations are `words', and transition rules are `grammar'. & \cite{smith1971formal}; \cite{culik1989limit,lee2003simulation,zhisong2005complexity,delacourt2017characterisation} \\
        \ $D$-spectrum & The number of distinct densities occurring in the periodic part of the CA's phase space trajectory. & \cite{fates2003experimental} \\
        \ R\'enyi entropy & Spatial, temporal or local parameterised generalisation of Shannon entropy on the space of CA configurations. & \cite{renyi1961measures}; \cite{Wuensche1999,Li1990}  \\
        \ Computational complexity & Judges whether computational problems are in principle solvable, and classifies CAs according to the predictability of their rule. & \cite{sutner1989computational}; \cite{green1987np,sutner2002cellular,sutner2012computational} \\ \bottomrule
    \end{tabular}
    \label{tab:list-of-math-tools-genotypic}
\end{sidewaystable}

\begin{sidewaystable}
    \centering
    \caption{Mathematical tools and concepts used for characterising and classifying CAs based on the phenotype (model outcome). The first reference indicates the original academic work. Selection largely copied from Vispoel et al.~\cite{Vispoel2022}.}
    \begin{tabular}{L{.25\linewidth}p{.6\linewidth}L{.15\linewidth}} 
        \textbf{Tool/concept} & \textbf{Brief description} & \textbf{References} \\ \toprule
        \multicolumn{3}{l}{\textsc{Behavioural classes}} \\
        \ Wolfram classes & Identifies ECA phenotype classes I (homogeneous), II (periodic), III (chaotic), IV (complex).  & \cite{Wolfram1984}; \cite{Langton1990} \\
        \ Li-Packard classes & Characterises ECA phenotype as null, fixed point, periodic, locally chaotic, chaotic. & \cite{Li1990}; \cite{Li1992}\\
        \ Culik-Yu classes & Formal definition of Wolfram classes. & \cite{Culik1988} \\ \midrule
        \multicolumn{3}{l}{\textsc{Local properties}} \\
        \ Mean-field theory & Measure of time-dependent configuration averages, most notably mean state density. & \cite{Wolfram1983}; \cite{Schulman1978} \\
        \ Local structure theory & Generalisation of mean-field theory to properties of blocks of cells. & \cite{Gutowitz1987}; \cite{Gutowitz1990} \\
        \ Two-point correlation & Measure for spatial correlations between local configuration structures. & \cite{Wolfram1983} \\
        \ Lyapunov exponents & Quantifies the sensitivity to initial conditions based on analysis of the difference pattern (damage spreading). & \cite{shereshevsky1992lyapunov,bagnoli1992damage}; \cite{Li1990transition,ninagawa1998fluctuation,bagnoli1999synchronization,tisseur2000lyapunov,courbage2006lyapunov,baetens2011topological,baetens2018lyapunov} \\
        \ Kolmogorov complexity & Time-dependent entropy-related measure of configuration complexity. & \cite{Kolmogorov1963}; \cite{zenil2010compression,zenil2013asymptotic,ninagawa2014classifying} \\ 
        \ Spectral properties & Investigates the CA frequency spectrum via some well-defined notion of a discrete Fourier transform. & \cite{ninagawa2008power}; \cite{ninagawa1998fluctuation,andrecut2000noise,ruivo2016computing}\\ \bottomrule
    \end{tabular}
    \label{tab:list-of-math-tools-phenotypic}
\end{sidewaystable}

We will use some of these properties as tools to distinguish between various CA families and types, either by demonstrating that the properties differ with those of their elementary counterpart, or by arguing that the properties are ill-defined for the non-classical CA at hand.

\subsection{Overview of CA selection and presentation of our methodology}
\begin{figure}
    \centering
    \includegraphics[width=\linewidth]{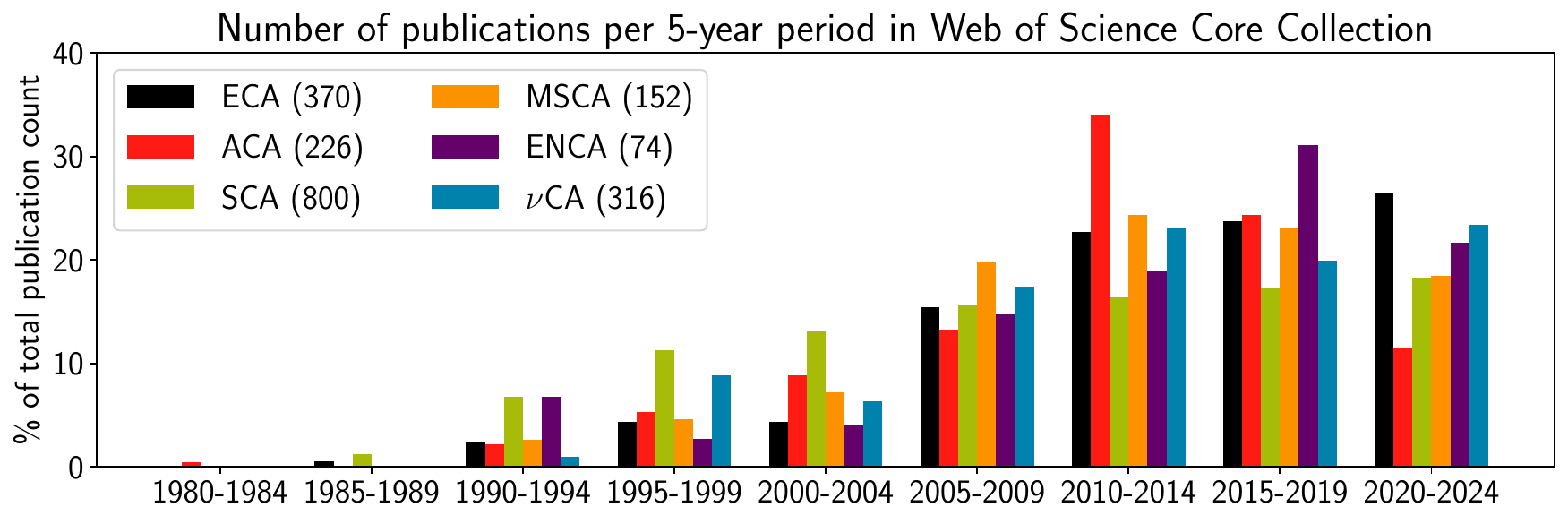}
    \caption{The number of publications found in the Web of Science Core Catalogue from searching six different topics, grouped in five year periods, and normalised over the total number of publications. These topics are ``elementary cellular automaton/a'' (black), ``asynchronous cellular automaton/a'' (red), ``stochastic/probabilistic cellular automaton/a'' (green), ``continuous/fuzzy cellular automaton/a'' (orange), ``network automaton/a'' \textsc{or} ``graph/nonlocal cellular automaton/a'' (purple), and ``non-uniform/hybrid cellular automaton/a''. The absolute number of recorded publications is indicated in parentheses in the legend. The colour coding is identical to that in Tab.~\ref{tab:list-of-ca-research}, giving an idea of which CA family is typically used for which application.}
    \label{fig:wos_publications}
\end{figure}

All five CA families that (generally) require an extension of the classical definition are listed in Tab.~\ref{tab:nECAs}. For completeness we highlight that this table contains CAs for which some non-trivial version of the definition is still within the domain of the classical CA (e.g.~a $k$-state CA). On the other hand, some families are omitted, either because we consider them both rare and inert (e.g.~`polygeneous' CAs \cite{burks1970essays}), or because they are too remote from the canonical research field and would deserve a separate survey (e.g.~quantum CAs \cite{arrighi2019quantum}).
For a historical perspective, the bar chart in Fig.~\ref{fig:wos_publications} shows the number of publications per five-year period, associated to ECAs and to each of the five CA families. The absolute numbers in this figure are of course highly dependent on the formulation of the query in the search engine of Web of Science (cf.~figure caption), but an interesting trend is visible in the normalised counts. First, an increasing number of articles is written for most CA families, with the clear exception of asynchronous CAs. Moreover, the number of new papers on non-uniform CAs and continuous/fuzzy CAs seems to have reached a plateau.

\begin{table}[h]
    \centering
    \caption{List of CA families discussed in this review paper. An asterisk ($^*$) indicates that such a complete survey is missing, followed by cited papers that perform a \textit{partial} review.}
    \begin{tabular}{L{.2\linewidth}p{.54\linewidth}L{.16\linewidth}}
        \textbf{CA family} & \textbf{Nature of the definition extension} & \textbf{Surveys}\\ \toprule
        Asynchronous (ACA) & The local update rule $\phi$ no longer applies to all cells simultaneously. & \cite{schonfisch1999synchronous}, \cite{bandini2012}, \cite{fates2014}\\
        Stochastic (SCA) & Cells update their state (partially) randomly with regards to location, timing, or update rule. & $^*$\cite{arrighi2013stochastic}, \cite{mairesse2014around}, \cite{louis2018probabilistic}, \cite{roy2022temporally} \\
        Multi-state (MSCA) & The state space is no longer binary but $k$-nary or continuous. & $^*$\cite{adamatzky1994hierarchy}, \cite{mingarelli2006study}, \cite{baetens2014towards} \\
        Extended neighbourhood (ENCA) & The neighbourhood $\mathcal{N}(c_i)$ of a cell $c_i$ is larger in space and/or time, differently shaped, or no longer consist of its direct (`physical') neighbours. & $^*$\cite{marr2009outer}, \cite{alonsosanz2009memory}, \cite{zaitsev2017generalized} \\
        Non-uniform ($\nu$CA) & The local rule $\phi$ can take more arguments than simply the neighbourhood configuration, allowing for heterogeneity. & \cite{dennunzio2012nonuniform}, \cite{bhattacharjee2020survey} \\ \bottomrule
    \end{tabular}
    \label{tab:nECAs}
\end{table}

We follow the same approach within each of the five families of non-classical CAs. First, we motivate the family's existence and formalise its definition. Second, we present the taxonomy of the various CA (sub)types belonging to the family at hand, generally guided by the kind and the degree of the extension. We highlight when particular definitions encompass others within the same family, and especially when they are mathematically equivalent across families. Third, we demonstrate how the genotype and phenotype of these CAs compare to that of ECAs and other CA families. We indicate when a particular characteristic (referring to Tabs.~\ref{tab:list-of-math-tools-phenotypic} and \ref{tab:list-of-math-tools-genotypic}) breaks down, inhibiting a quantitative comparison. Fourth, we briefly discuss how the unique character of the CA family allows for a number of applications (referring to Tab.~\ref{tab:list-of-ca-research}).

\begin{figure}
    \centering
    \includegraphics[width=.5\linewidth]{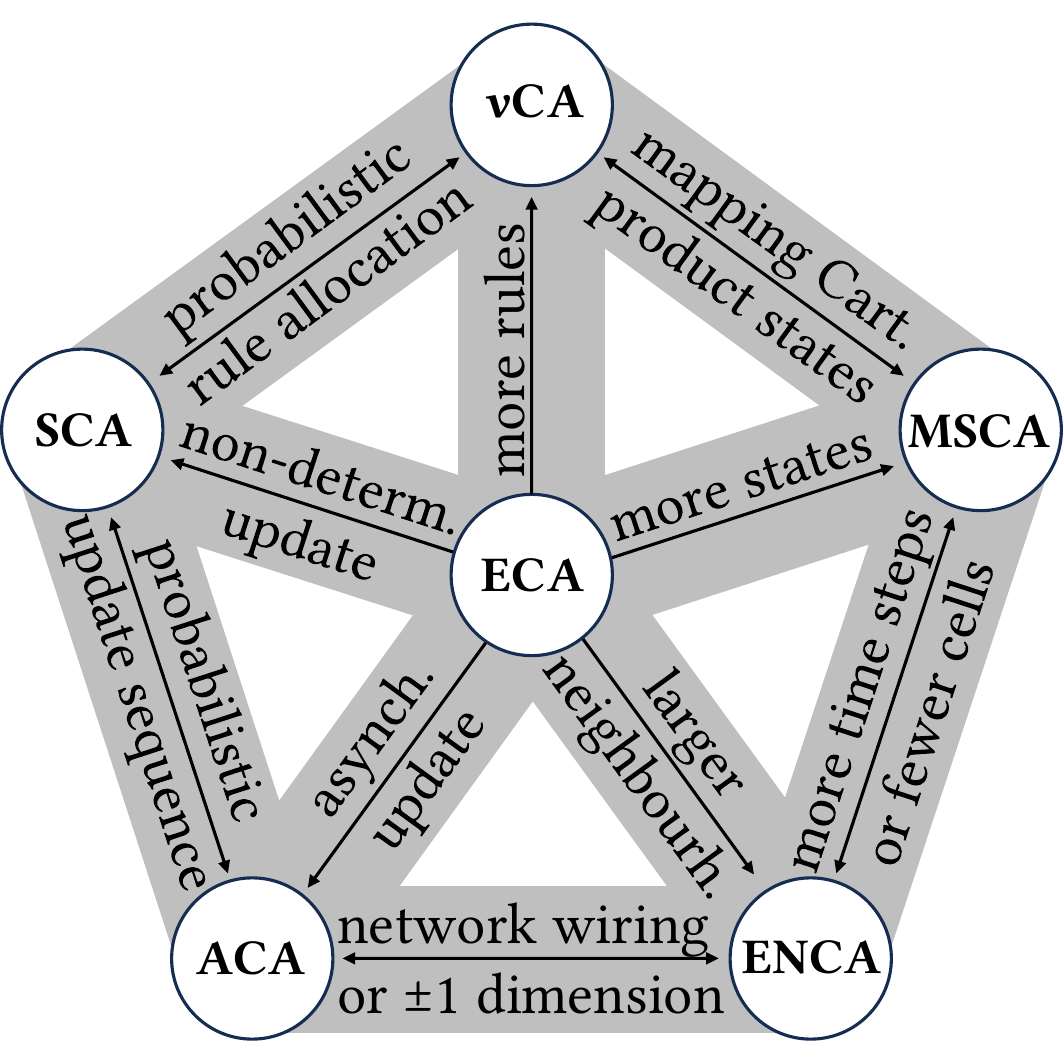}
    \caption{We identify five families of CAs, based on their divergence from the elementary cellular automaton (ECA). Some important identities and trade-offs between these families are indicated by arrows annotated with the required change in interpretation. Clockwise, the diagram contains non-uniform ($\nu$CA), multiple-state (MSCA), extended-neighbourhood (ENCA), asynchronous (ACA), and stochastic cellular automata (SCA).}
    \label{fig:CA-family-diagram}
\end{figure}

Fig.~\ref{fig:CA-family-diagram} shows the central diagram related to the taxonomy presented in this review paper. This diagram summarises the relationship of the various CA families with the ECA and with each other, and will serve as a convenient mnemonic device for the reader while we progress through our detailed review.

\section{Asynchronous CAs}
\label{sec:asynchronous_CA}

\subsection{Motivation and definition}

Global synchronisation of the local update rule is contradictory to the CA philosophy (claiming to be `fully local'), and often does not reflect the nature of the modelled phenomenon; nature has no `central clock' \cite{ruxton1998realism}. For these reasons, the assumption of synchrony was challenged by Priese \cite{priese1978note} (investigating the mathematical and computer-scientific implications), and Ingerson and Buvel \cite{ingerson1984structure} (with a more phenomenological approach).

Of course synchrony \textit{does} occur in nature (see e.g.~references in \cite{Ravignani2017}), but rather than interpreting this observation as a justification for top-down synchrony-by-design, it better suits the spirit of CA research to regard this as a challenge to find bottom-up synchrony emerging from the dynamical system. In fact, it was counter-intuitively shown in the recent past that global synchronisation (from arbitrary initial conditions) can even \textit{only} be achieved by going beyond synchronous deterministic updating \cite{richard2017synchronisation}. In any case, this suggests that the study of asynchronous CAs (ACAs) is not merely an exercise in computational tinkering, but motivated by the principles of a proper scientific model.

Expanding on the ECA paradigm, we define elementary ACAs by expanding Def.~\ref{def:ECA} (see also Fig.~\ref{fig:CA-family-diagram}).
\begin{definition}[Asynchronous cellular automaton]
\label{def:ACA}
An asynchronous cellular automaton (ACA) is identical to an ECA (Def.~\ref{subsec:elementary-cas}), with the exception of the sixth property, which becomes:
\begin{itemize}
    \item[6.] The dynamics of cell $c_i$ at time $t$ is determined by \textit{either} the identity rule (the state remains unchanged), \textit{or} one of the possible elementary Wolfram rules $\phi : \{0,1\}^3 \rightarrow \{0,1\}$.
\end{itemize}    
\end{definition}
\noindent Spatiotemporal variations in this choice define the various ACA subtypes.

\subsection{ACA taxonomy}
\label{subsec:aca_taxonomy}

Despite the long tradition of ACAs, no overall consensus exists in literature on what defines a particular subtype, nor on its nomenclature. Relatively recent work concerned with statistics on ACA spacetime diagrams (e.g.~\cite{fates2014,bandini2012,schonfisch1999synchronous}), has however somewhat converged on a number of subtypes, of which we display the taxonomy below.

\subsubsection{Spatial ACA variations}
\paragraph{Partially asynchronous CAs}
An `$\alpha$-asynchronous CA', also known as an $m$-ACA \cite{dennunzio2013mACA} or block-sequential ACA \cite{aracena2009robustness}, allows for partial synchrony, and is contrasted with what Fat\`es calls \textit{fully asynchronous updating} \cite{fates2014}. Whether or not a cell is updated depends on the `synchrony rate' $\alpha \in [0,1]$. The limit situation where $\alpha=1$ reduces the ACA to the synchronous case, while $\alpha=0$ induces an entirely static CA.

The synchrony rate is mostly (and most generally) interpreted as a probability on a cell-per-cell basis:
\begin{align}
    s(c_i, t+1) = \begin{cases} \phi(\tilde{s}(\mathcal{N}(c_i), t)), & \text{with probability } \alpha,\\
    s(c_i, t), & \text{with probability } 1-\alpha.
    \label{eq:nECA_alpha-asynchronism0}
    \end{cases}
\end{align}
We identify the set of updated cell indices at time step $t$ as $Z(t) \subseteq \{1, \ldots, N\}$. Since $\vert Z(t)\vert \sim \text{Bin}(N,\alpha)$, with expectation value $\alpha N$, Eq.~\eqref{eq:nECA_alpha-asynchronism0} is (up to a rounding error) equivalent with
\begin{align}
    s(c_i, t+1) = \begin{cases} \phi(\tilde{s}(\mathcal{N}(c_i), t)), & \text{if } i \in Z(t),\\
    s(c_i, t), & \text{else}.
    \end{cases}
    \label{eq:nECA_alpha-asynchronism}
\end{align}

Clearly, there are many other ways to choose the set $Z(t)$, resulting in different dynamics. Worsch \cite{worsch2013towards}, for example, discusses a particular type called a `neighbourhood-independent' ACA, additionally demanding that neighbouring cells cannot update simultaneously. It is more common practice, however, to demand that the number of updated cells per time step is always \textit{exactly} $\lfloor\alpha N\rfloor$. The fixed-size subset $Z(t)$ is typically chosen in an entirely stochastic fashion, allowing for $\dbinom{N}{\vert Z(t)\vert}$ choices. Note that if $Z(t)=Z$ is static, we arrive at a trivial two-rule spatially non-uniform CA, where one of the rules is the identity operator (see Section \ref{sec:non-uniform_CA}).

Some authors have explored updating different cells with different fixed periods or phases (e.g.~Ingerson and Buvel \cite{ingerson1984structure}) which is equivalent to \textit{deterministically} selecting $Z(t)$ at any time step. Note, however, how in these cases one tyrant overthrows another: the global deterministic clock is replaced by a global deterministic selection of the update set. Such a non-stochastic $Z(t)$ is interesting from a mathematical point of view, however, as Worsch \cite{worsch2013towards} demonstrated that for all deterministic choices of $Z(t)$, a single ACA with a radius-1 rule can simulate all others. That is to say: precisely the same dynamics are achieved by either (i) choosing a set $Z(t)$ given a rule $\phi$, or (ii) choosing a rule $\phi$ given a set $Z(t)$. Similar equivalences were shown by Aracena et al.~\cite{aracena2009robustness} for random Boolean networks (see Section \ref{sec:non-classical-neighbourhood_CA}) with various asynchronous update schemes. In a classic paper, Golze \cite{golze1978simulating} showed how to construct a synchronous CA in $d+1$ dimensions from an ACA in $d$ dimensions, by reinterpreting time as a spatial dimension.

The ACA is \textit{fully} asynchronous if $\vert Z(t) \vert = 1$. Many authors, including Sch\"onfisch and De Roos \cite{schonfisch1999synchronous}, consider \textit{full} synchrony to be more natural in the context of mathematical modelling, because physical time can be infinitely divided (up to the Planck limit). Synchrony, in other words, is a fictional notion, as events are never \textit{really} simultaneous.

\paragraph{Fully asynchronous CAs}

The most general approach to fully asynchronous CAs, a `random independent' ACA from a uniform choice, is to select and update a single cell at random, i.e.
\begin{align}
    s(c_i, t+1) = \begin{cases} \phi(\tilde{s}(\mathcal{N}(c_i), t)), & \text{if } i = k,\\
    s(c_i, t), & \text{else},
    \end{cases}
    \label{eq:nECA_random-independent}
\end{align}
where $k$ is chosen randomly and independently from the set $\{1, \ldots, N\}$. The number of updates of cell $i$ in $N$ time steps is again binomially distributed $\sim \text{Bin}(N, 1/N)$ with expectation value 1. Next, for `randomly ordered' ACAs with a `random new sweep', we further impose that updating the $N$ cells occurs in cycles. That is to say: every $N$-th time step, a permutation of the tuple $(1, \ldots, N)$ is randomly chosen, and the cells are updated in this order. This prevents cells from updating more than once every cycle, and makes sure any given cell is updated at least once after $2N$ time steps. Narrowing this down further, said permutation is chosen randomly but only once (`random fixed sweep'), resulting in any given cell being updated after exactly $N$ time steps. One may distinguish further the particular case where the updating happens according to ascending cell number, in a `line-by-line sweep'. This assures, after all, that always precisely one neighbour will have been just updated, resulting in spatial correlations \cite{rajewsky1997exact}. In any of these fully-asynchronous cases, the expectation value for the number of time steps between two consecutive updates is simply $N$ (see Tab.~1 in \cite{schonfisch1999synchronous} for all statistics).

\subsubsection{Temporal ACA variations}

Sch\"onfisch and De Roos distinguish `step-driven' and `time-driven' asynchronous updating \cite{schonfisch1999synchronous}. In the former approach -- tacitly assumed up to this point -- some rhythmic notion of discrete time remains. In the latter approach, discrete time is abandoned and the update of a cell depends on some continuous time-dependent probability, typically according to an exponential distribution. The difference between CAs with various \textit{temporal} asynchronies is arbitrary, however. After all, due to the unidirectionality of time, a bijective mapping between step-driven updates (computer time) and time-driven updates (physical time) is always possible \cite{fates2014}.

\subsection{ACA genotype and phenotype}

Introducing asynchronous updating affects the way CAs can be described mathematically, as well as the way they behave dynamically.

\subsubsection{The ACA genotype}

All metrics used to classify the rule table (Tab.~\ref{tab:list-of-math-tools-genotypic}, top) are still fully applicable to any of the above ACA types, with respect to the rule $\phi$ that updates the selected ACA cells. The rule table itself could be extended to include, with a particular probability, the identity operator; this is however identical to the rule table of a stochastic CA, and will therefore be discussed in Section \ref{sec:stochastic_CA}.

Global properties (Tab.~\ref{tab:list-of-math-tools-genotypic}, bottom) are affected not only by the local rule $\phi$, but also by the selected subset $Z(t)$. Generally, therefore, the mathematics regarding attractor structure and topological dynamics now involve $Z(t)$, hindering closed expressions, but some progress was made for particular cases. Hansson et al.~\cite{hansson2005asynchronous} formally identified all 1-D ACAs with periodic boundary conditions (including non-elementary CAs) that are independent of the asynchronous update order. Amongst these ACAs, Macauley and Mortveit \cite{macauley2013atlas} reviewed and catalogued those with a random fixed sweep for all 104 ECA rules for which the set of periodic points is independent of update sequence. Manzoni \cite{manzoni2012asynchronous} generalised a number of mathematical properties (e.g.~surjectivity and injectivity) to study periodic orbits in the space of configurations of fully-asynchronous CAs, highlighting how these formal results can contribute to expanding knowledge of properties that are already known for synchronous CAs.

Regarding computational complexity, some ACAs have been shown to support universality \cite{adachi2004computation}. While more recent universal ACAs \cite{fei2021effect} are less complex than the original proposals, they however still require non-uniformity of the local update rule (see Section \ref{sec:non-uniform_CA}).

\subsubsection{The ACA phenotype}

Spacetime diagrams of ACAs can be analysed in exactly the same fashion as those of ECAs, keeping in mind that fewer cell updates occur during a fixed time interval. For a fair comparison, ACA diagrams are therefore typically cropped to only show configurations after (on average) all cell states are updated. Behavioural classifications and local properties (Tab.~\ref{tab:list-of-math-tools-phenotypic}) can subsequently be deduced from the resulting diagram, e.g.~entropy-based classification \cite{lei2021entropy}. Notably, the choice of $Z(t)$ introduces additional degrees of freedom that can be perturbed when investigating the CA phenotype, especially when results are plotted against the quasi-continuous $\alpha$ parameter. Fat\`es \cite{fates2014} emphasises that this is of particular help in the (computational) study of self-organisation and robustness.

Baetens et al.~\cite{baetens2012effect} use Lyapunov exponents \cite{bagnoli1992damage,bagnoli1999synchronization} to examine the stability of 2-D totalistic ACAs, comparing four fully-asynchronous update schemes. Using mean-field theory based on a large number of simulations, they show that introducing asynchronicity strongly affects stability and may even change the CA's Wolfram class. This is confirmed in the study by Bour\'e et al.~\cite{boure2012probing}, who additionally suggest that the robustness properties of a system may actually be used to probe the desired update scheme of the underlying mathematical model. Lumer and Nicolis had done similar research some ten years prior \cite{lumer1994synchronous} on 2-D \textit{continuous} CAs (or `coupled-map lattices', see Section \ref{sec:multi-state_CA}). Using tools directly related to the Lyapunov exponents, they demonstrated that ACAs remain stable for higher values of the logistic map's nonlinearity parameter, compared to their synchronous counterpart, especially for a random new sweep. Perhaps most convincing in its simplicity, however, is the study by Bandini et al.~\cite{bandini2012}. These authors show that non-trivial phenotypical change arises from some type of full asynchrony for 9 out of the 16 one-neighbourhood CAs, despite being even simpler than ECAs.

\subsection{Applications of ACAs}

Next to being of strictly methodological interest, ACAs are applied in the study of parallel computation (e.g.~\cite{huberman1993evolutionary}), and for a wide range of mathematical models \cite{ruxton1996effects,gunji1990pigment,bezbradica2014comparative,bersini1994asynchrony}. Applications are generally motivated by the ACA's demonstrated enriched phenotype, and by a sense of realism (no central clock).

\subsubsection{Computational advances}

Faced with practical considerations, most notably the scaling of semiconductor elements and the demand for long-range interconnection \cite{monica2016cellular}, early research into nano-scale computers had shifted from Von Neumann's theoretical automaton to simpler CAs by the early 90s \cite{biafore1994cellular}. Peper et al.~\cite{peper2003laying} showed how not only ACAs support integration into electrical circuits, but how asynchronicity actually allows for the construction of `delay-insensitive' circuits \cite{keller1974towards}. The relatively recent work by Fei et al.~\cite{fei2021effect} discusses a 3-state, 10-rule ACA that uses its asynchronicity for the 2-D crossing of signals. Asynchronicity in these computational units therefore protects the system from inherent noise \cite{gharavi1992effect}, and is required for efficient information transfer.

Beyond hardware implementations, asynchronicity has its merits in solving computational tasks such as the prisoner's dilemma \cite{huberman1993evolutionary}, where asynchronicity is sometimes even \textit{required} for the existence of a solution. Ruivo et al.~\cite{ruivo2020} discuss a 4-neighbour ACA that solves the global synchronisation problem -- which is impossible for synchronous CAs \cite{richard2017synchronisation}. Asynchronous solutions have also been put forward to tackle the density problem \cite{tomassini2001evolving} and the parity problem \cite{ruivo2019perfect}.

\subsubsection{ACAs in mathematical modelling}

Asynchronicity has been used for CA-based modelling in every domain listed in Tab.~\ref{tab:list-of-ca-research}. Most notably, ACAs are used in biology and chemistry, where the loss of synchrony is almost invariably either a way of including stochasticity, or a pragmatic choice for (speedy) computation. In biology, for example, Messinger et al.~\cite{messinger2007task} use partial asynchrony for investigating biology-inspired task-performing dynamics, and Sieburg et al.~\cite{sieburg1990hiv} use the fully asynchronous variation to simulate HIV infection in an artificial immune system. In chemical modelling, Perez-Brokate et al.~\cite{perezbrokate2014overview} reviewed  `blocked' $\alpha$-asynchronism to model corrosion, Seybold et al.~\cite{seybold1997kinetics} used stochastic partial asynchronicity to simulate chemical kinetics, and Kier et al.~\cite{kier2000cellular} reviewed similar simulations for ACAs with a random new sweep. Full asynchronicity with a random order is also imposed onto Van der Wee\"en et al.'s CA model for photocatalytic degradation \cite{vanderweeen2012modeling}. We stress, however, that none of these examples use `pure' ACAs, but rather CAs with more than one extension, typically including multiple states and an extended neighbourhood (cf.~Sections \ref{sec:multi-state_CA} and \ref{sec:non-classical-neighbourhood_CA}).

More interesting than listing how all ACA types are used in various domains of science, is to highlight a number of applications in mathematical modelling where the asynchronicity \textit{itself} was studied. Ruxton \cite{ruxton1996effects} studied the effect of four update types in a very simple ecological model (see also Ref.~\cite{hogeweg1988eco}) of stochastic and temporally non-uniform ACAs. He shows how this choice strongly affects the average time until extinction, and generalises these results in later work \cite{ruxton1998realism}. Another early biological work, by Gunji~\cite{gunji1990pigment}, shows how macroscopic patterns on mollusc can be generated by various ACA types, indicating that the particular update scheme influences the emergent patterns even more than the choice of the local update rule. Bezbradica et al.~\cite{bezbradica2014comparative} studied the effect of various updating schemes in the context of molecular pharmacology, and found that a random independent ACA hosts the most robust models of structural chemical interactions. Such stabilising effects had already been accentuated a decade earlier by Bersini and Detours \cite{bersini1994asynchrony} for an abstract immune network model. Similar effects are observed for stochastic CAs, which we will investigate in the next section. We observe how this is largely the consequence of the high similarity between both definitions of the CA extensions.

\section{Stochastic CAs}
\label{sec:stochastic_CA}

\subsection{Motivation and definition}

Most natural processes appear to exhibit some degree of unpredictability, and randomness has been long identified as a driving force for many types of structure formation, e.g.~morphogenesis \cite{turing1990morphogenesis}. Adding stochasticity to CAs therefore again expresses realism in mathematical modelling \cite{ruxton1998realism}. From a methodological point of view, a non-deterministic setup also significantly affects a CA's emergent behaviour, e.g.~in terms of sensitive dependence on initial conditions \cite{Baetens2017}. The motivation for stochastic CAs (SCAs, also `probabilistic' CAs \cite{louis2018probabilistic}) is therefore very similar to that for partially asynchronous CAs, and so is its definition; most generally, we again alter the sixth property in Def.~\ref{def:ECA} (cf.~Fig.~\ref{fig:CA-family-diagram}).
\begin{definition}[Stochastic cellular automaton]
\label{def:SCA}
A stochastic cellular automaton (SCA) is identical to an ECA (Def.~\ref{subsec:elementary-cas}), with the exception of the sixth property:
\begin{itemize}
    \item[6.] The dynamics of cell $c_i$ at time step $t$ is governed by
    \begin{align}
        s(c_i, t+1) = \begin{cases} \phi(\tilde{s}(\mathcal{N}(c_i), t)), & \text{with probability } p,\\
        \psi(\tilde{s}(\mathcal{N}(c_i), t)), & \text{with probability } 1-p.
        \end{cases}
        \label{eq:SCA_definition}
    \end{align}
\end{itemize}
\end{definition}
\noindent Typically $p$ is independent of time and space, clearly reducing to the elementary case if $p=1$. Time and/or space dependence is however often desirable with applications in mind; these options generate the landscape of SCA types discussed next.

\subsection{SCA taxonomy}

We first observe some notable special cases. When $\psi$ is the identity rule in Eq.~\eqref{eq:SCA_definition}, the SCA reduces to an $\alpha$-asynchronous CA (cf.~Eq.~\eqref{eq:nECA_alpha-asynchronism0}), with $\alpha = p$. We also observe that if rules $\phi$ and $\psi$ are not complementary, i.e.~when $W(\phi) + W(\psi) \neq 255$, the CA is only partially stochastic, as some neighbourhood configurations will generate an identical outcome for both $\phi$ and $\psi$. At the opposite side of the stochastic spectrum we find SCAs with complementary `null' rules $\phi$ and $\psi = \phi^\text{C}$ with $W(\phi) = 255 - W(\phi^\text{C}) = 0$. This induces complete neighbourhood independence and loss of information, which for $p=1/2$ clearly leads to pure noise. Due to its simplicity and its intuitive link to sociological behaviour, we additionally highlight the `majority SCA' \cite{slowinski2015phase}, which is equivalent to choosing $W(\phi)=232$ and $W(\psi)=51$ in Eq.~\eqref{eq:SCA_definition}.

\subsubsection{Spatial SCA variations}

\paragraph{Local space dependence}

The state transition of a cell is determined by the states of its neighbouring cells in the classical sense, but may also be influenced by this neighbourhood in a \textit{stochastic} (yet local) sense. Observe a particular SCA where the probability $p$ is also neighbourhood-dependent:
\begin{align}
    s(c_i, t+1) = \begin{cases} 1, & \text{with probability } p(\tilde{s}(\mathcal{N}(c_i))),\\
    0, & \text{with probability } 1-p(\tilde{s}(\mathcal{N}(c_i))).
    \end{cases}
    \label{eq:SCA_definition2}
\end{align}
Because for ECAs there are $k^{|\mathcal{N}|} = 8$ different neighbourhood configurations, such an SCA is entirely defined by the octuple $\mathbf{p} = (p(0,0,0), \ldots, p(1,1,1)) = (p_0, \ldots, p_7) \in [0,1]^8$. Note how this matches the definition in Eq.~\eqref{eq:SCA_definition}, if we demand that $\psi = \phi^\text{C}$, and if we perform a suitable permutation of $\mathbf{p}$. Clearly, many choices for $\mathbf{p}$ are possible; we highlight three that induce some structure between the tuple's elements.

First, diploid CAs are a type of neighbourhood-dependent SCAs proposed by Fat\`es \cite{Fates2017}, and are motivated by a formal investigation into symmetry breaking in biological systems resulting from inherent stochasticity. This particular type of SCA is based on the definition in Eq.~\eqref{eq:SCA_definition2}, but reduces the dimensionality of $\mathbf{p}$ from eight to three by setting
\begin{align}
    p(\tilde{s}(\mathcal{N}(c_i))) = \lambda \phi(\tilde{s}(\mathcal{N}(c_i))) + (1-\lambda) \psi(\tilde{s}(\mathcal{N}(c_i))).
\end{align}
For a non-trivial diploid CA, we have that $\phi \neq \psi$ and $0 < \lambda < 1$, such that (again referring to Eq.~\eqref{eq:SCA_definition2}) $p \in \{0, \lambda, 1-\lambda, 1\}$, or $p \in \{\lambda, 1-\lambda\}$ if $\psi = \phi^\text{C}$. For a particular $\lambda \in (0, 1)$ there are 8088 pairs $\{\phi, \psi\}$ that result in a unique diploid rule \cite{Fates2017}.

Second, Mairesse and Marcovici \cite{mairesse2014around} discuss some particular cases of two-state, $|\mathcal{N}|=2$ SCAs that also rely on a reduction of the probability tuple dimensionality. These cases are entirely defined by $k^{|\mathcal{N}|} = 4$ probabilities in the quadruple $\mathbf{p} = (p_0, p_1, p_2, p_3)$. Clearly, such CAs are of very little interest for applications, but they do invite exhaustive mathematical treatment. Four special one-parameter cases are the `noisy additive SCA' with parameters $(0, p, p, 0)$, the `symmetric noisy additive SCA' with parameters $(1-p, p, p, 1-p)$, the `Stavskaya SCA' with parameters $(0, p, p, p)$, and the `directed-animals SCA' with parameters $(p, 0, 0, 0)$.

Third, a well-studied case is the Domany-Kinzel CA \cite{domany1984equivalence}. This CA is totalistic, minimally asychronous -- odd (even) cells are only updated at odd (even) time steps --, and determined by the probability octuple $\mathbf{p} = (x, y, x, y, y, z, y, z)$. This model's demonstrated equivalence to an Ising model is of particular interest to percolation theory in statistical physics \cite{domany1984equivalence}.

\paragraph{Global space dependence}

If the probability $p = p(c_i)$ depends on the position of the cell, the SCA is globally space-dependent and cannot be considered uniform. After all, two transition rules with different associated probabilities may be considered as distinct rules altogether. We investigate spatially non-uniform CAs in Section \ref{sec:non-uniform_CA}. Here, we simply note that many applications require a position-dependent probability due to some global parameter, e.g.~temperature in Vichniac's Ising model \cite{vichniac1984} (see Section \ref{subsec:applications-of-scas}).

\subsubsection{Temporal SCA variations}
\label{subsubsec:temporal-SCA-variations}

Most generally, the probability may again depend on some global model parameter, which is now time-dependent, such that $p = p(t)$. This breaks uniformity, but now in a temporal sense, and is again hard to map due to its highly idiosyncratic ties to the model at hand.

What \textit{can} be investigated methodologically is what is known as a `temporally stochastic CA' \cite{roy2022temporally}. Here all cells also follow either rule $\phi$ or rule $\psi$ in a spatially uniform sense, but the global parameter that decides this assignment is stochastic:
\begin{align}
    \left(s(c_i, t+1)\right)_{i=1}^N = \begin{cases}
        \Phi(\left(s(c_i, t)\right)_{i=1}^N), & \text{with probability } 1-\pi,\\
        \Psi(\left(s(c_i, t)\right)_{i=1}^N), & \text{with probability } \pi.
    \end{cases}
\end{align}
Recall that $\Phi : S^N \rightarrow S^N$ (resp.~$\Psi$) is the global transition rule induced by the local transition rule $\phi$ (resp.~$\psi$). Probability $\pi$ is called the `temporal noise rate', as a sporadic global update according to $\Psi$ is often considered a transient noisy `hiccup' in the system. Note that this again fits into the general definition of SCAs (Def.~\ref{def:SCA}) upon setting
\begin{align*}
    p(t) = \begin{cases}
        1, & \text{with probability } 1-\pi,\\
        0, & \text{with probability } \pi.
    \end{cases}
\end{align*}

\subsection{SCA genotype and phenotype}

Introducing stochasticity generates a number of interesting differences and subtleties regarding both input and output of the dynamical system, as compared to ECAs. Below we discuss the SCA genotype and phenotype respectively, again referring to Tabs.~\ref{tab:list-of-math-tools-phenotypic} and \ref{tab:list-of-math-tools-genotypic} for background information.

\subsubsection{The SCA genotype}

In terms of the rule table parameters, a naive extension for SCAs consists of applying the definition to either of the applying rules $\phi$ or $\psi$, or to a probability-weighted average of both. From Def.~\ref{def:SCA}, the mean-field rule table parameters, and hence the Langton parameter, may for example be defined as
\begin{align*}
    M_i = \sum_{\substack{\mathcal{N}\text{ with } i \\\text{non-}q\text{ values}}}\biggl(p[\phi(\mathcal{N}) \neq q] + (1-p)[\psi(\mathcal{N}) \neq q]\biggr),
\end{align*}
where the square brackets output 0 or 1 depending on the truth condition, and $q$ is the quiescent state. Note how this definition naturally allows for summing over a neighbourhood-dependent probability $p = p(\mathcal{N})$. Despite the elegance of these parameter extensions, it is not a priori evident (and in general untrue \cite{roy2022temporally}) that the properties of the SCA update rule are `somewhere in between' of those of its constituent rules. An extension rooted in a weighted average is therefore not necessarily useful. Investigating correlations between particular SCA rule-table property definitions and their actual behaviour is an open direction of further inquiry.

A different, global approach arises from the observation that consecutive SCA configurations form a Markov chain on the state space $S^N \rightarrow S^N$. This allows for adopting the Markov chain formalism in the study of CA attractor structure, mapping and analysing the distribution of future configurations. The main topics of interest are (i) short-term time-dependent behaviour, (ii) long-term steady-state behaviour, and (iii) the study of the `absorption time' \cite{parzen1999stochastic}, to which some theoretical insight has been provided \cite{busic2013probabilistic, agapie2014probabilistic}. A recurring theme is to determine whether or not a particular SCA is ergodic, i.e.~whether it will have visited all $2^N$ configurations for $t\rightarrow\infty$. Maes and Shlosman provided a sequence of general criteria for ergodicity in SCAs \cite{maes1991ergodicity}. Particular SCAs may however require a more tailor-made approach, e.g.~Holroyd et al.'s recent proof of ergodicity related to a `percolation game' \cite{holroyd2019percolation}.

\subsubsection{The SCA phenotype}

An analysis of SCA spacetime diagrams is generally mathematically identical to that of ECAs. It may, however, reveal distinct classifications or local properties due to inherent randomness.

Regarding classification, first note that no SCA can be considered \textit{truly} homogeneous or periodic, as randomness will disturb any repetitive pattern in consecutive configurations. Additionally, while Wolfram's class IV (complexity) may still be a pertinent classification visually, the capacity to support universal computation can no longer be an argument for assigning this class. Indeed, Arrighi et al.~showed that universal SCAs do not exist \cite{arrighi2013stochastic}. Roy et al.~\cite{roy2022temporally} exhaustively investigated the classification of temporally stochastic CAs, for all \num{3825} combinations of non-equivalent Wolfram rules. They show i.a.~that (i) two rules that are classified as chaotic, may join forces in an SCA to generate periodicity (and vice versa), and that (ii) a behavioural phase transition may occur for particular values of the temporal noise rate. We note that the quantitative results of their research are contingent on what constitutes a `class', which is mathematically insufficiently defined for SCAs, and for CAs in general \cite{Culik1988}. This nonetheless does not affect the validity of the qualitative observations, demonstrating paradoxically that randomness can aid stability.

Regarding local properties, the phenomenology of an SCA can be investigated in function of the probability $p$, often demonstrating `jumps' in behaviour when $p$ surpasses a particular value. Such a jump may indicate a phase transition or a system bifurcation; Bagnoli and Rechtman \cite{bagnoli2018phase} provide an overview of different types of possible phase transitions in SCAs. More specifically, Baetens et al.~\cite{Baetens2017} mapped such dynamics of SCAs by means of a sensitivity measure, the maximum Lyapunov exponent, and the Lempel-Ziv complexity, for a large number of simulations of SCAs. Their research is limited to SCAs with $\psi$ the identity operator, such that the results apply to $\alpha$-asynchronous CAs as well. They demonstrate that an SCA's sensitivity to initial conditions and its tendency towards chaos is highly dependent on the update probability; a small change in $p$ may radically alter the system's behaviour. Similar results have been achieved for diploid SCAs where $\psi$ is not the identity rule. A highly-cited study by Martins et al.~\cite{martins1991evidence} shows that this CA can accommodate three phase transitions. The third phase was discovered by studying the influence of slightly changing initial conditions -- i.e.~studying ``damage spreading'' \cite{bagnoli1996damage}. This clearly has important consequences for the physical systems that this CA models, such as reaction-diffusion systems. Fat\`es \cite{Fates2017} demonstrates second-order phase transitions and symmetry breaking when varying the mixing parameter $\lambda$, by mean-field analysis of (kinks) density. Cirillo et al.~\cite{Cirillo2021} studies this in more detail (stochastic `reset'), and provides some theoretical justifications of Fat\`es's numerical results for mixtures with the null rule. Also majority-voter SCAs exhibit such behaviour, both in one \cite{busic2013probabilistic}, and two dimensions \cite{slowinski2015phase}.

\subsection{Applications of SCAs}
\label{subsec:applications-of-scas}

Above-mentioned methodological advances have inspired applications in computer science and mathematical modelling (and vice versa) \cite{louis2018probabilistic}. Systems that are robust to noise -- or even require noise -- are especially valuable in the former, while the phenomenological richness provided by SCA phase transitions is highly relevant for the latter.

\subsubsection{Computational advances}

Robustness to noise is a highly desirable property of nanoscale computing systems, so a lot of research has been done on CAs that can `remember information' \cite{gacs2001reliable}. Specific computational tasks may also be solved by SCAs, most notably the density classification problem (e.g.~\cite{fuks2002nondeterministic}).

\subsubsection{SCAs in mathematical modelling}

A classical SCA application is Vichniac's implementation of the Ising model for ferromagnetism \cite{vichniac1984}. He shows that an approach with deterministic CAs leads to a ``feedback catastrophe'', which not at all represents the observed physical behaviour of magnetic metals. This is solved by associating a temperature-dependent transition probability to each of the possible neighbourhood configurations, and is therefore another curious example of a system that \textit{requires} randomness to achieve stability. Research related to the Domany-Kinzel CA \cite{domany1984equivalence,martins1991evidence} further solidifies this relationship with statistical physics.

As Baetens et al.~\cite{Baetens2017} conclude, an SCA's propensity for phase transitions necessitates scrutiny while developing a discrete mathematical model. Additionally, the inclusion of stochastic parameters of course generates a much wider spectrum of conceivable model setups. While sometimes an exhaustive grid search is feasible (e.g.~\cite{VanderWeeen2011}), genetic algorithms exist to help determining which stochastic rule may be most pertinent to describe (and predict) the physical data at hand (e.g.~\cite{billings2003identification}). The latter approach was recently used for parameter optimisation for an SCA-based epidemiological model for COVID-19 \cite{ghosh2020covid}, acquiring quite remarkable results. The latter model (and many other non-toy models) also includes extended neighbourhoods (Section \ref{sec:non-classical-neighbourhood_CA}) and multiple states, to which we turn next.

\section{Multi-state CAs}
\label{sec:multi-state_CA}

\subsection{Motivation and definition}

The general definition of Von Neumann (cf.~Def.~\ref{def:CA_von-neumann}) allows for more than two states, but the ECA definition (Def.~\ref{def:ECA}) should be generalised to accommodate more than two states (cf.~Fig.~\ref{fig:CA-family-diagram}).
\begin{definition}[Multi-state cellular automaton]
    A multi-state cellular automaton (MSCA) is identical to an ECA (Def.~\ref{subsec:elementary-cas}), with the exception of the second property:
\begin{enumerate}
    \item[2.] $S$ is the state set, which may or may not be finite, and which may or may not have some structure imposed on it.
\end{enumerate}
\end{definition}
\noindent MSCAs serve as a basis for a wide variety of applications \cite{reichenbach2008self,GANGULY2002,sirakoulis2000cellular,zanette1992multistate,Alexandridis2011,nishinari2000multi} because it is rarely the case that a natural system allows only two discrete states. Still, an in-depth understanding of the dynamics of such models and a robust mathematical framework is mostly lacking.

Most of the theoretical work on CAs has been done for two-state CAs. This is in part due to the observation that for some (mostly computational) objectives, the study of binary-state CAs is rich enough by itself. It was shown, for example, that replicating machines in the framework of CAs could also be established in a two-state CA \cite{berlekamp1982winning}. It is also due to the overwhelming complexity of the mathematics when allowing more than two states: considering three rather than two states in a 1-D CA with $|\mathcal{N}|=3$ increases the number of possible local update rules from 256 to $\sim10^{13}$. While this can be an advantage, e.g.~for random number generation \cite{Bhattacharjee2017pseudo}, it often obstructs theoretical progress.

MSCAs with a continuous state space are called continuous CAs (CCAs). Rucker raises a number of objections against the study of CCAs \cite{Rucker2003}, the most important of which is redundancy. CCAs are similar to well-established finite-difference methods for numerically solving differential equations \cite{Wolfram2002}, and also similar to coupled-map lattices \cite{garcia-morales2016deterministic}. Rucker also emphasises that the CA approach is different, however, being more rooted in computational experiment and observation, rather than theory and proof.

\subsection{MSCA taxonomy}

Below, we first distinguish between ternary, $k$-nary, and continuous-valued CAs, zooming in on a particular type of continuous-valued CA called `fuzzy' CAs. Note that we no longer distinguish between spatial and temporal variations, as variations only affect the state space and its internal structures.

\paragraph{Ternary and $k$-nary CAs}

\begin{figure}
    \centering
    \includegraphics[width=.5\linewidth]{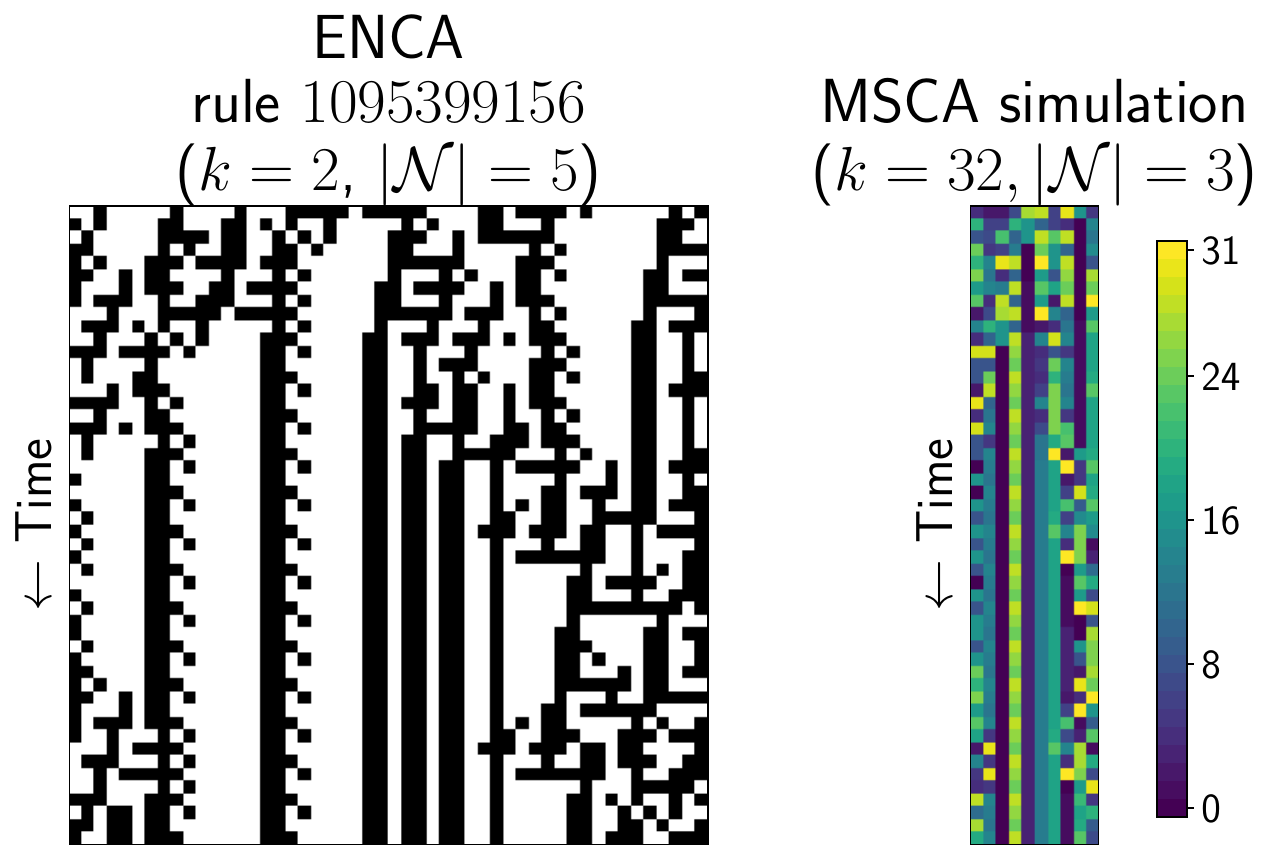}
    \caption{Complexity trade-off between neighbourhood size, number of cells, and number of states. Here a $k=2, |\mathcal{N}|=5$ CA with 50 cells is simulated by a $k=32, |\mathcal{N}|=3$ CA with 10 cells, without loss of information in either direction.}
    \label{fig:smith-tradeoff}
\end{figure}

Generally, for $k$-nary CAs we find $k^{k^{\vert\mathcal{N}\vert}}$ possible local update rules. It is shown that a CA with a local rule defined for a particular neighbourhood is equivalent to a CA with neighbourhood size three and a different state space \cite{smith1971cellular}, by `pooling' cells together (see Fig.~\ref{fig:smith-tradeoff}). This implies that only considering the $|\mathcal{N}|=3$ case is sufficiently general.

Following the convention for ECAs, a ternary CA rule can be identified by associating the number $s_i \in \{0, 1, 2\}$ with each of the 27 possible neighbourhoods. This may in turn be translated to a decimal number, i.e.
\begin{align*}
    (s_{26}\ldots s_0)_3 = \left(\sum_{i=0}^{26} 3^{i}s_i\right)_{10},
\end{align*}
where the subscript indicates the numeral base. Due to the quickly increasing number of possible neighbourhood configurations for increasing $k$, often only totalistic $k$-nary CAs with $S = \{0, \ldots, k-1\}$ are considered (see e.g.~\cite{Wolfram2002}). This reduces the number of rules (or ``codes'') to $k^{(k-1)\vert\mathcal{N}\vert + 1}$.

\paragraph{Continuous-valued CAs}

A $k$-nary CA in the limit of $k \rightarrow \infty$, where the state space is typically mapped to $S = [0,1]$, can take an infinite number of values. If additionally $S$ is continuous, the entire object is called a continuous-valued (or simply continuous) CA (CCA).

A complete rule table would now be infinitely large. In order to nonetheless quantify transition rules, the state set $S$ is considered to be a field, and some (neighbourhood-dependent) function applies on its elements. In the language of coupled-map lattices, this is called a `map', and the `coupling' is achieved by including the neighbourhood. Wolfram showed that CCAs can also display the rich and often chaotic dynamics found in their discrete counterparts \cite{Wolfram2002}.

Referring back to Section \ref{sec:stochastic_CA}, we note that a neighbourhood-dependent SCA is a particular case of a CCA, provided we interpret the continuous values as probabilities for the cells to be either in state 0 or state 1. An SCA is entirely determined by the tuple $\mathbf{p} = (p_0, \ldots, p_7)$, such that the continuous local update rule becomes
\begin{multline*}
    \phi(s_{i-1}, s_i, s_{i+1}) = p_0\overline{s_{i-1}}\ \overline{s_{i}}\ \overline{s_{i+1}}
    + p_1\overline{s_{i-1}}\ \overline{s_{i}}\ s_{i+1}
    + p_2\overline{s_{i-1}} s_{i} \overline{s_{i+1}}
    + p_3\overline{s_{i-1}} s_{i} s_{i+1} \\
    + p_4s_{i-1}\overline{s_{i}}\ \overline{s_{i+1}}
    + p_5s_{i-1} \overline{s_{i}}s_{i+1}
    + p_6s_{i-1}s_{i} \overline{s_{i+1}}
    + p_7s_{i-1}s_{i}s_{i+1},
\end{multline*}
where $\overline{s_i} = 1-s_i$, and $s_i$ is shorthand notation for $P(s(c_i, t)=1)$. Other particular relationships can be inferred for specific types of SCAs, such as diploid SCAs.

\paragraph{Fuzzy CAs}

Fuzzification is understood as generalising Boolean-logical \textsc{and}, \textsc{or} and \textsc{not} operators such that they can operate on all elements in $S$, typically with the aim of expressing a `degree of truth' rather than simply true (1) or false (0) \cite{cattaneo1997fuzzy}. Fuzzy-state CAs (FCAs) constitute a particular type of MSCAs where the state space $S$ is the unit interval, but where the governing rules are defined from a binary-state CA. The FCA framework forms an elegant bridge between $k$-nary CAs and continuous CAs, which allows for a more well-defined analysis of chaotic dynamic behaviour.

The FCA local rule is defined as the `fuzzification' of the local rule of a binary CA; it is defined from a rule written as a Boolean expression (cf.~\cite{Wolfram2002} p.~884) in `disjunctive normal form'. Wolfram's rule $18$, for example, can be expressed as
\begin{align*}
    \phi(s_{i-1}, s_i, s_{i+1}) = (\overline{s_{i-1}} \land \overline{s_i} \land s_{i+1}) \lor (s_{i-1} \land \overline{s_i} \land \overline{s_{i+1}}),
\end{align*}
where $\land$ and $\lor$ denote the logical \textsc{and} and \textsc{or}, respectively, and an overbar denotes the \textsc{not} operator.

For a formal definition and a better application-driven methodology, we refer to Doostfatemeh and Kremer \cite{doostfatemeh2005new}. Note that CAs have also been called `fuzzy' when referring not to the states themselves, but rather to the choice of a deterministic transition rule applying to a non-continuous CA \cite{adamatzky1994hierarchy}. These CAs are however more akin to SCAs.

\subsection{MSCA genotype and phenotype}

\subsubsection{The MSCA genotype}

Most rule table parameters remain well-defined for $k$-nary CAs. Provided $q \in S$ is a quiescent state, the definition of the Langton parameter $\lambda$ still holds. As a matter of fact the parameter was originally introduced for examining the rule space of $\vert\mathcal{N}\vert \geq 5, k \geq 4$ CAs \cite{Langton1990}, and the paper's main results are only valid for rule spaces significantly larger than that of ECAs. The resulting $\lambda$ values are roughly correlated to the four Wolfram classes. The $Z$-reverse and obstruction parameters still hold for deterministic $k$-nary CAs. The $\mu$-sensitivity can also be computed, but some care should be taken in defining (or perhaps quantifying) a `change of state', as now more than one option is available.

Global properties of $k$-nary CAs are generally difficult or impossible to achieve or even define, in part again due to undecidability \cite{Culik1988}. More becomes possible when some structure is imposed on $S$ (e.g.~linearity allows for a study of attractor structures \cite{manzini1999attractors}) or when particular rules are investigated in detail (e.g.~rule 2460N in 2-D \cite{siap2011characterization}). Conversely, it is also possible to determine particular types of $k$-nary CA that are required to possess a specific property. For example, Bhattacharjee et al.~\cite{Bhattacharjee2016} pinpoint a number of \textit{reversible} $k$-state finite CAs. More recently, Wolnik et al.~\cite{Wolnik2020split} shared a formula for finding all number-conserving $k$-state CAs, and showed that the only \textit{reversible} number-conserving ternary CAs are shifts that are intrinsically 1-D \cite{Wolnik2020ternary}.\\

For fuzzy CA, one could of course simply adopt all rule table parameters associated with the ECA the fuzzy CA is defined from. Due to the definition of fuzzification, this should indeed reflect the FCA's properties when $\forall i: s_i \in \{0,1\}$, but it is not obvious how the rule table parameters are to be interpreted when $s_i \in\ ]0,1[$. Neither is it evident that such parameters would convey useful information; after all, FCAs and binary CAs can exhibit similarities \cite{betel2011fuzzy}, but fuzzy logic may also lead to entirely different behaviour. Flocchini et al.~\cite{flocchini2000convergence} show, for example, how a randomly initialised FCA based on rule 90 will evolve to a homogeneous value of $1/2$, while paradoxically its Boolean counterpart is entirely aperiodic \cite{jen1990aperiodicity}. For the characterisation of an FCA rule, and for all types of continuous CAs, it is therefore generally more insightful to adopt the formalism of coupled-map lattices, e.g.~for determining steady states (attractors). An overview thereof is provided by Kaneko \cite{kaneko1992overview} and more recently by Bunimovich \cite{bunimovich2005coupled}.

\subsubsection{The MSCA phenotype}

A classification of $k$-nary CAs according to emergent behaviour is generally still possible, and often even more systematic (cf.~Langton parameter). Nonetheless, no classification appears to be entirely deterministic, which is complicated even further by the large number of variations on periodicity or complexity with larger $k$.

Local properties require some attention in order to apply for $k$-nary CAs, but the finiteness of the state set generally enables a logical extension of the definition. Welch \cite{welch1984technique}, for example, extended the definition of Lempel-Ziv complexity to accommodate any symbol, in the context of data compression. For many other properties, however, the nature of the extension depends on whether the state set is additive or not. State density, for example, is no longer defined when states are mere symbols with no mathematical relationship, such as the $S$, $I$ and $R$ states in epidemiology. Additionally, as the state at the previous time step may be important for the measured property, a change of state is no longer unambiguous. This is highlighted in the recent publication by Vispoel et al.~\cite{vispoel2023lyapunov}, continuing work on the Lyapunov exponent for ternary CAs initiated by Baetens and De Baets \cite{baetens2014towards}.

Behaviour-based classification of CCAs is typically based on the attractor structure associated with the update function (map), i.e.~the location and type of fixed point(s). This behaviour generally changes (and bifurcations can occur) for different parameter values of the map, such as the $r$ parameter in the logistic map. For FCAs an exhaustive classification is provided by Mingarelli \cite{mingarelli2010classification}. Most mean-field parameters and correlation functions can easily be extended for CCAs, but measures for sensitivity to initial conditions and complexity are again generally better expressed using the tools and methods developed for coupled-map lattices.

\subsection{Applications of MSCAs}

\subsubsection{Computational advances}

Applications of MSCAs within the domain of computer science are as old as the entire research field, as Von Neumann's universal constructor \cite{neumann1966theory} required 29 states. Computational tasks are generally more difficult to solve, e.g.~the density problem for ternary CAs \cite{fuks2020explorations}. The additional complexity of adding more states can however also be exploited, such as in random number generation \cite{Bhattacharjee2017pseudo}.

\subsubsection{MSCAs in mathematical modelling}

Some successful work on the edge of theory and application has been done, including a multi-state extension of the Ising model \cite{wu1982potts}, the Greenberg-Hastings model for excitable media \cite{greenberg1978spatial}, and studies in artificial life in three-state CAs \cite{wuensche2005glider}.

Examples of $k$-nary CA applications for mathematical modelling are of course widespread, because realism almost invariably requires more than two states. Examples include biological competition \cite{reichenbach2008self}, pattern classification \cite{GANGULY2002,Das2009}, epidemiology \cite{lu2023spatial,sirakoulis2000cellular}, immunology \cite{da2005dynamic}, reaction-diffusion \cite{zanette1992multistate}, forest fire modelling \cite{Alexandridis2011}, stock market dynamics \cite{bartolozzi2004}, and traffic flow modelling \cite{nishinari2000multi}.

FCAs have been applied in the context of pattern recognition \cite{maji2005fuzzy}, but also in artificial life research \cite{reiter2002fuzzy}, forest fire modelling \cite{mraz2000fuzzy}, or for modelling snowflakes on a hexagonal grid \cite{coxe2003fuzzy}. As indicated, dynamics of more general CCAs essentially overlap with finite difference methods for partial differential equations, and are especially useful for modelling physics \cite{toffoli1984alternative}. Some notable examples of CCAs used in this fashion are a CA model for diffuse and dissipative systems \cite{chan1995cellular}, the reproduction of the Navier-Stokes equation in theoretical hydrodynamics \cite{wolfram1986fluids}, a model for groundwater dynamics \cite{ravazzani2011macroscopic}, and a model for seismic elastodynamics \cite{leamy2008application}. Conversely, PDEs can simulate CAs with complex behaviour \cite{omohundro1984modelling}.

\section{CAs with extended neighbourhoods}
\label{sec:non-classical-neighbourhood_CA}

\subsection{Motivation and definition}
\label{subsec:ENCA_def}

A CA as defined in Defs.~\ref{def:CA_von-neumann} and \ref{def:ECA} is characterised by local and uniform interaction: a cell only interacts with its neighbours, and the definition of this neighbourhood $\mathcal{N}$ is identical relative to any cell $c_i$ in the tessellation $\mathcal{T}$. This is narrowed further for ECAs (Def.~\ref{def:ECA}), where $\mathcal{T}$ is a tessellation of $\mathbb{R}$ and $\mathcal{N}(c_i) = \{c_{i-1}, c_i, c_{i+1}\}$ is limited to its direct neighbours in one dimension.

This neighbourhood definition is of course quite restrictive. From a mathematical point of view, it is of interest to study the effect of altering the neighbourhood size and regularity on the dynamical behaviour. As a neighbourhood of maximal size is identical to a global system, varying neighbourhood parameters allows for the examination and quantification of `quasi-local' systems. Moreover, from the perspective of mathematical modelling, relaxing the neighbourhood condition is nearly always indispensable. First, in terms of dimensionality: 3-D events can often be reduced to 2-D models by omitting the least influential dimension (in a top-view fashion), but losing another dimension often means losing important information. Second, in terms of neighbourhood size: many phenomena cannot realistically be assumed to only and fully rely on its closest neighbours, and even if they would, these neighbours are rarely distributed in a perfect grid, such that an approach more akin to mean-field theory presents itself. Third, `local' interactions can be physically and temporally separated, and may very well differ for distinct elements in the model. This is an important difference between e.g.~a lattice of magnetic dipole moments in the Ising model, and an irregular grid (network) of friends on the Internet. Both the methodological and the application-driven perspective on this family of extended CAs motivate further inquiry.

A so-called `adjacency matrix' $\mathbf{A}$ contains all neighbourhood information, and consists of  elements $A_{ij}$ that are non-zero if cell $c_i$ contains cell $c_j$ in its (non-local) neighbourhood. If $\forall i,j: A_{ij} \in \{0,1\}$, this allows for $2^{N^2}$ possibilities. For ECAs with periodic BCs, the adjacency matrix alone is sufficient for expressing all neighbourhoods since $\mathbf{A}$ is tridiagonal. 

Other interesting CA families are identified with particular choices for $\mathbf{A}$, and are discussed next. They are all spatial or temporal variations of the following definition extension (cf.~Fig.~\ref{fig:CA-family-diagram}):
\begin{definition}[Extended-neighbourhood cellular automaton]
A finite (size-$N$) extended-neighbourhood cellular automaton (ENCA) is a sextuple $\mathcal{C} = \langle \mathcal{T}, S, s, s_0, \mathbf{A}, \phi\rangle$, defined as in Def.~\ref{def:ECA} with the exception of properties 5 and 6:
\begin{enumerate}
    \item[5. ] The adjacency matrix $\mathbf{A}$ is implicitly defined as $$\mathcal{N}(c_i) = \left\{ c_j \vert A_{ij} \neq 0\right\}_{j=1}^N,$$
    where $A_{ij}$ is an element of $\mathbf{A}$ and $\mathcal{N}(c_i)$ is the neighbourhood of cell $c_i$.
    \item[6. ] The function $\phi : S^{|\mathcal{N}|(t+1)} \rightarrow S$ determines the next state of cell $c_i$, i.e.~$$s(c_i, t+1) = \phi(\tilde{s}(\mathcal{N}(c_i),0), \tilde{s}(\mathcal{N}(c_i),1), \ldots, \tilde{s}(\mathcal{N}(c_i),t)),$$ where $\tilde{s}(\mathcal{N}(c_i),t)$ is the tuple $\left(s(c_{j},t) | A_{ij}(t) \neq 0 \right)^{N}_{j=1}$. This function $\phi$ is called the `local update rule'.
\end{enumerate}
\end{definition}

\subsection{ENCA taxonomy}

\subsubsection{Spatial ENCA variations}

\paragraph{Larger local neighbourhood}

First, consider enlarging the neighbourhood size symmetrically (although some authors discuss asymmetric neighbourhoods as well \cite{bandini2012}) around the cell with a radius $r \in \mathbb{N}$ such that $\vert \mathcal{N} \vert = 2r + 1$. The elements of the adjacency matrix then become
\begin{align}
A_{ij} = \delta_{ij} + \sum_{k=1}^r(\delta_{(i+k)j} + \delta_{i(j+k)}),
\end{align}
where $\delta_{ij}$ is the Kronecker delta, with indices mod $N$. The number of possible local update rules now increases to $k^{k^{2r+1}}$. More systematic approaches to an enlarged neighbourhood are sometimes considered to enable a mathematical description and a link to other CA families, e.g.~\cite{Kayama2016}.

\paragraph{Higher dimensions and irregular tessellations}

Second, virtually all applications require CAs with more than one spatial dimension. In a regular square grid with $N^2$ cells, two symmetrical neighbourhoods are typically considered, called the Von Neumann neighbourhood and the Moore neighbourhood. The former consists of the central cell and all directly adjacent cells, while the latter additionally contains all diagonally adjacent cells. Mathematically, these neighbourhoods are the simplest non-trivial form of `templates' $H_r^{(n)}$ and $J_r^{(n)}$, first defined by Cole \cite{cole1969real}:
\begin{align*}
    H_r^{(n)} = \left\{ \mathbf{a}\ \biggl|\ \sum_{i=0}^{n-1} |a_i| \leq r \right\}, \qquad
    J_r^{(n)} = \left\{ \mathbf{a}\ \biggl|\ \max\limits_{0\leq i \leq n-1}\{|a_i|\} \leq r \right\}.
\end{align*}
Here, $n$ is the CA dimension, and $\mathbf{a}$ is the vector indicating the location of neighbourhood cells, relative to the central cell. Note that $H_r^{(1)}=J_r^{(1)}$. Butler \cite{butler1974note} demonstrated that (in two dimensions) any CA with Moore neighbourhood of radius $r$ can be simulated by a CA with a larger state space and Von Neumann neighbourhood of radius $r'>r$, but performing calculations $r'/r$ times faster. These definitions can be extended to more general shapes and into higher dimensions (e.g.~\cite{baetens2012cellular}), but the fact remains that any $n$-D CA can be reduced to a 1-D CA with an extended neighbourhood.

The classical CA paradigm is built on the assumption that the tessellation $\mathcal{T}$ is defined over a regular domain. Yamada and Amoroso \cite{yamada1969tessellation} showed that any CA on a regular tessellation can be simulated by a CA on a different regular tessellation, provided we alter the state space. The most well-researched alternative is the hexagonal grid \cite{bays2009triangular}. CAs defined on irregular tessellations were examined by Baetens and De Baets \cite{baetens2010a}. Examples include a Voronoi-like tessellation \cite{adamatzky1996voronoi} and Penrose tilings \cite{owens2010investigations}. Margenstern \cite{margenstern2013hyperbolic} has gone even further, by considering CAs defined on hyperbolic space.

\paragraph{Network automata}

Third, neighbourhoods can be so irregular and/or non-local that it no longer makes sense to consider a CA as `cellular'. All cells (now typically called `nodes') are connected in a so-called `network automaton' (NA, also: `graph CA' or `nonlocal CA') \cite{tomita2002graph}, where the connectivity of nodes may again be represented in the adjacency matrix $\mathbf{A}$. The neighbourhood size $|\mathcal{N}(c_i)|$ now generally depends on the node, and in the context of network theory it is often referred to as the `node degree'. In general, the value $A_{ij}$ representing the connection from node $c_i$ to $c_j$ may be any positive real number, where typically a higher value indicates a `stronger' connection.

This of course allows for a virtually endless structural variation of networks, whose properties are studied in the field of `complex networks' \cite{mata2020complex}. When other NA properties such as state space and local update rule are still sufficiently reminiscent of the classical CA paradigm, e.g.~for random Boolean networks \cite{gershenson2012guiding}, theoretical knowledge and practical applications may nonetheless be transferred between both research domains.

\subsubsection{Temporal ENCA variations}

The adjacency matrix $\mathbf{A}$ may be time-dependent, such that neighbourhoods can vary as the CA evolves. Such CAs were suggested by Ilachinski \cite{ilachinski2009structurally} and coined `structurally dynamic' CAs. This dynamic structuring of neighbourhoods may be random, preserve some statistical property, be dependent on the state of the node, etcetera \cite{albert2000topology}. It may even allow for adding or removing nodes, which is studied in percolation theory \cite{li2021percolation}. This again opens Pandora's box of possible variations, and takes us beyond the scope of this CA-centred review paper. Within the confines of this review, however, we discuss the static temporal ENCA types named `memory CAs', and the related concept of `reversible' CAs.

\paragraph{Memory CA}

An ECA is considered to be memoryless in the sense that the new state of a cell depends on the neighbourhood configuration at the preceding time step only. This can however be extended by considering past states in the evolution of the CA, which particularly Alonso-Sanz has intensively researched (e.g.~\cite{alonsosanz2003elementary}). Mathematically, the local update rule in Def.~\ref{def:ECA} is extended such that $s(c_i,t+1) = \phi(m(\mathcal{N}(c_i),t))$, where
$$m(\mathcal{N}(c_i),t) = \biggl(\tilde{s}(\mathcal{N}(c_i),0),\allowbreak\tilde{s}(\mathcal{N}(c_i),1),\allowbreak\ldots,\allowbreak\tilde{s}(\mathcal{N}(c_i),t)\biggr).$$
Many ways of incorporating neighbourhoods in previous time steps present themselves, some of which support the design of desirable (complex) behaviour \cite{Martinez2013}. One particular ENCA type is the `average memory CA' \cite{alonsosanz2009memory}, where at every time step $t$, and for each cell, one computes a weighted mean of all states since the initial configuration. More often, however, the local rule incorporates a limited memory in the state transition \cite{Adamatzky2009}; it is said to be of order $M$ if it depends on $M$ previous time steps. Motivated by physical realism, a variation of memory CAs where some cells have a longer memory than others -- i.e.~non-uniform memory CAs, see Section \ref{sec:non-uniform_CA} -- has also been explored \cite{roy2019study}.

\paragraph{Reversible CA}

CAs are generally not `reversible', meaning that a particular configuration may have multiple predecessors: only six of the 256 ECAs are reversible, indeed \cite{Wolfram2002}, and all are trivial. As temporal reversibility (or technically, CPT invariance) is a fundamental symmetry in physics, CA models that aim to describe physics ideally reflect this property \cite{margolus1984physics}. Additionally, reversibility is desirable from a computer architecture perspective \cite{toffoli1984alternative}.

\begin{figure}[!htpb]
    \centering
    \includegraphics[height=.16\textwidth]{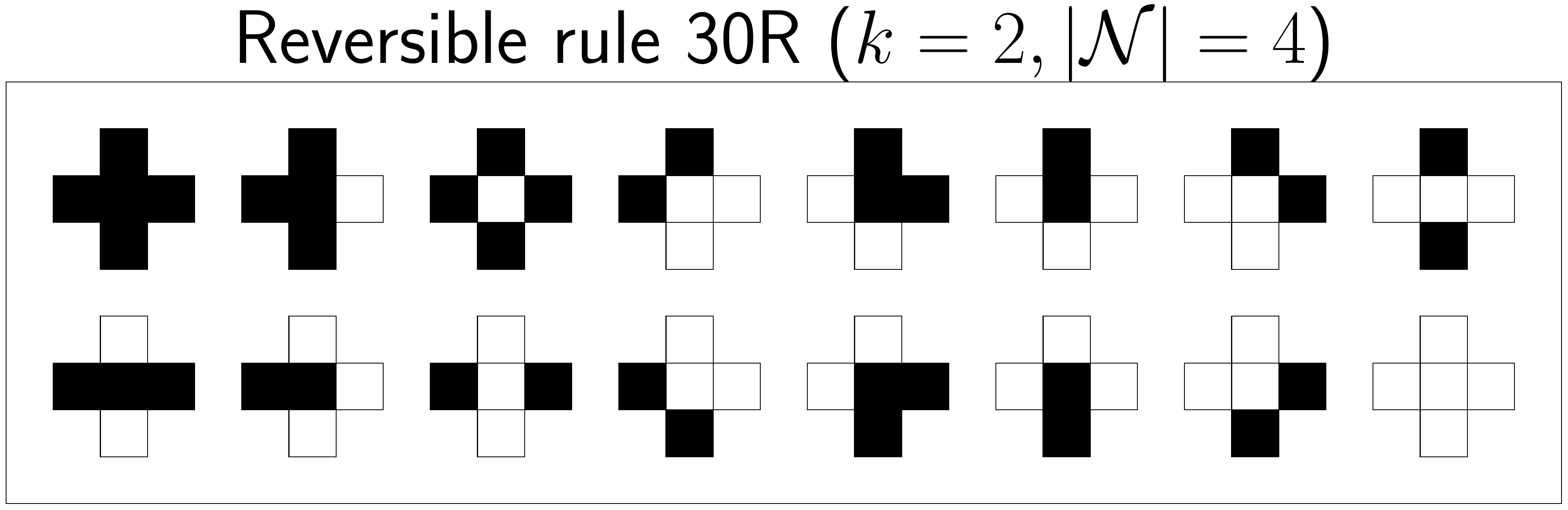} \quad
    \includegraphics[height=.16\textwidth]{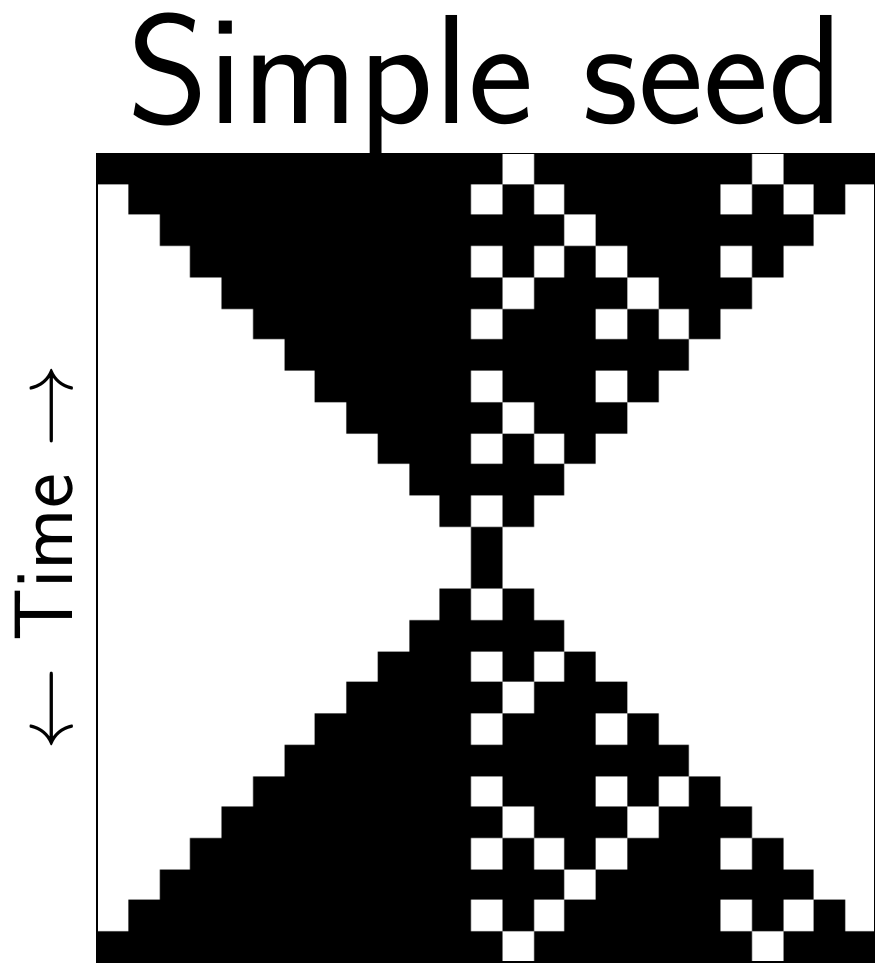} \quad
    \includegraphics[height=.16\textwidth]{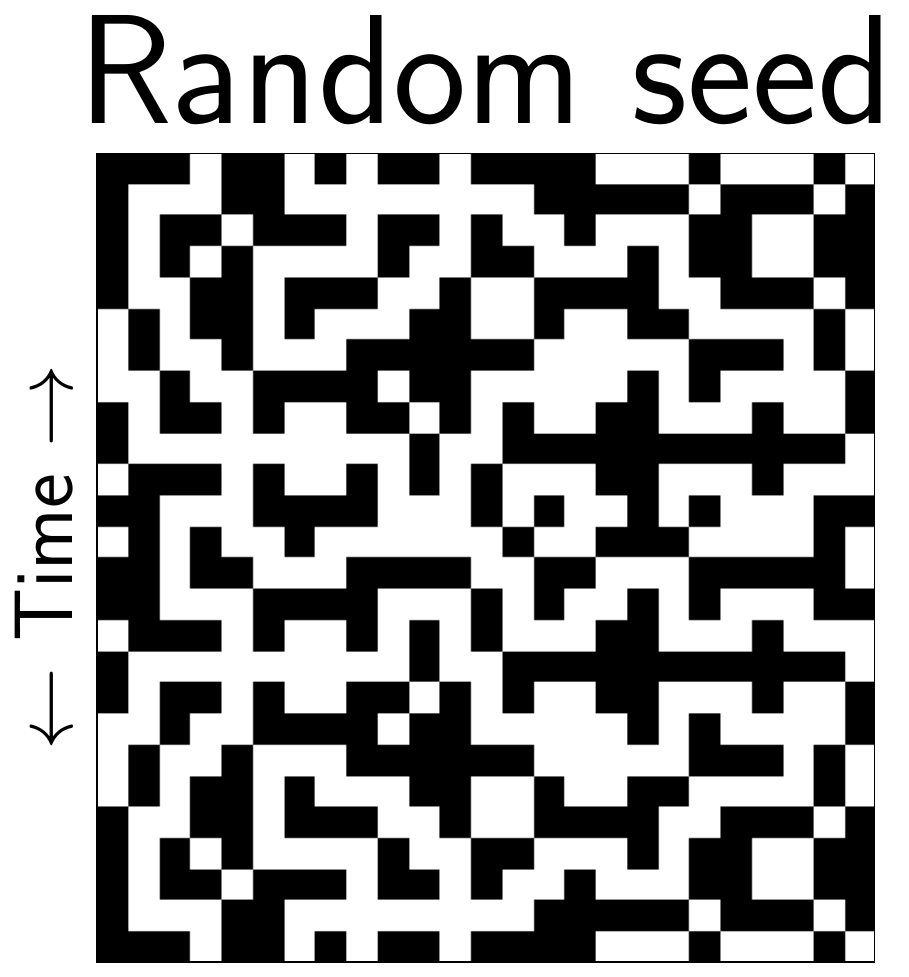}
    \caption{Rule 30 can be made reversible by imposing the complementary rule (rule 225) if the updated cell was in state 1 two time steps ago. The rule table is on the left, and two resulting spacetime diagrams, which are clearly symmetric under time reversal, are on the right.}
    \label{fig:rca_rules}
\end{figure}

Reversibility can be induced in ECAs by involving second-order memory ($M=2$) \cite{margolus1984physics}. The new state of a cell is inverted if two steps previously the cell was in state $1$, i.e.
\begin{align*}
    s(c_i,t+1) = \begin{cases}
        \phi(\tilde{s}(\mathcal{N}(c_i),t)), & \text{if } s(c_i,t-1)=0, \\
        1-\phi(\tilde{s}(\mathcal{N}(c_i),t)), & \text{if } s(c_i,t-1)=1.
    \end{cases}
\end{align*}
The resulting rule is simply called $W(\phi)$ with a suffix `R'; some examples of 1-D reversible CAs and their respective rule tables are shown in Fig.~\ref{fig:rca_rules}. Similar procedures have been investigated for 2-D CAs \cite{alonsosanz2003reversible}, and generalised for reversible CAs supported by higher-order memory \cite{secktuohmora2012invertible}. As an important remark in the context of our taxonomic treatment, Toffoli \cite{Toffoli1977computation} showed that any $n$-D CA can be simulated by a reversible CA of dimension $n+1$.

\subsection{ENCA genotype and phenotype}

The family of ENCAs contains a relatively wide variety of types. We discuss the genotype and phenotype of these types in the same order as these were presented above, and note that they generally exhibit less characteristic overlap than types in other CA families.

\subsubsection{The ENCA genotype}

\begin{figure}
    \centering
    \includegraphics[width=.7\linewidth]{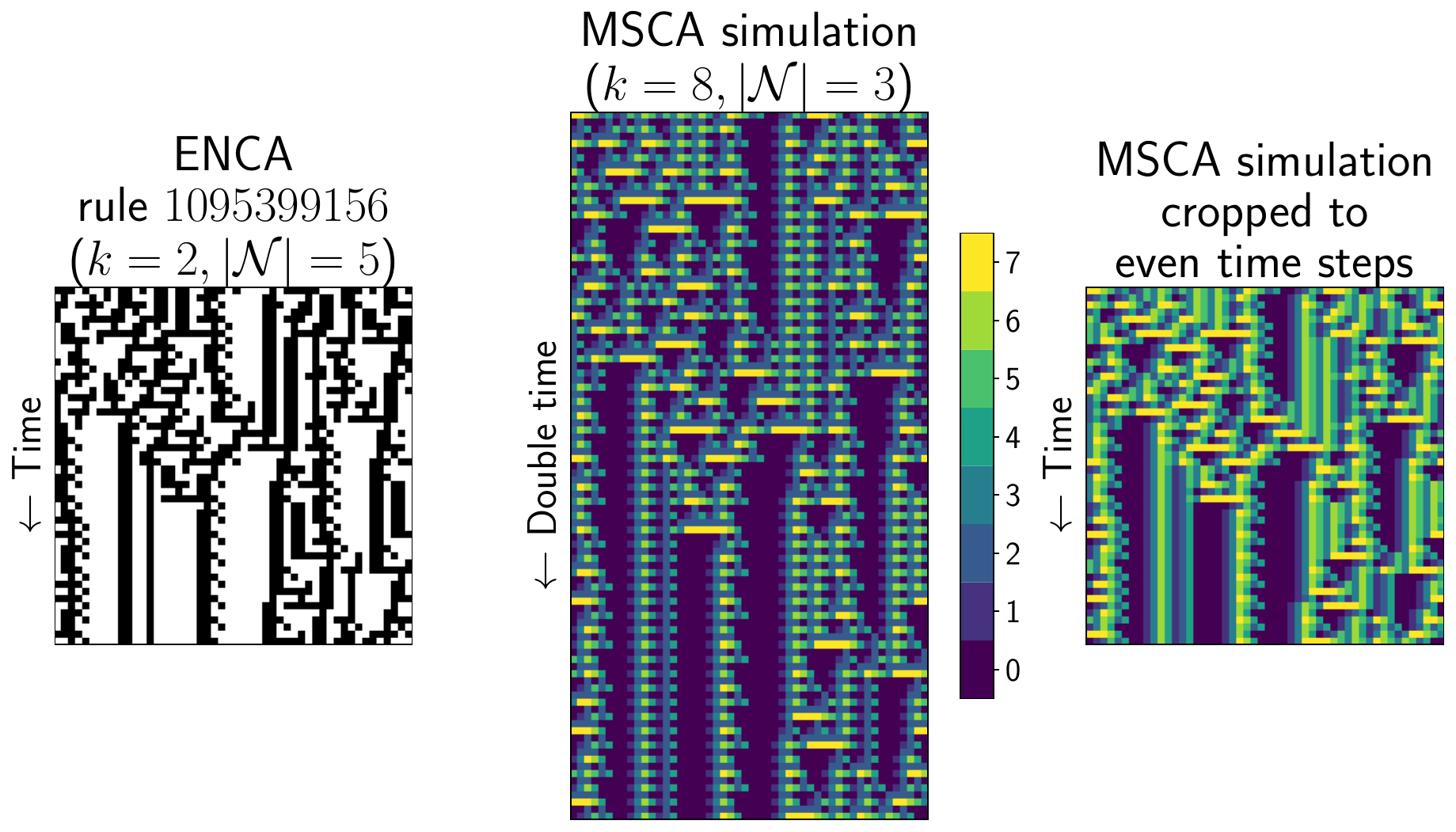}
    \caption{Complexity trade-off between neighbourhood size, number of time steps, and number of states. Here a $k=2$, $|\mathcal{N}|=5$ CA is simulated by a $k=8, |\mathcal{N}|=3$ CA with the same number of cells, but twice the number of time steps.}
    \label{fig:complexity-tradeoff2}
\end{figure}

When the neighbourhood size increases, this affects the rule table parameter values, as many more neighbourhood configurations make up the rule table. The definition of these parameters however remains identical (cf.~Tab.~\ref{tab:list-of-math-tools-genotypic}), again demonstrated by the fact that e.g.~the Langton parameter was originally investigated for $|\mathcal{N}|=5$ \cite{Langton1990}. Global properties are also still well-defined for larger neighbourhoods, but can become much more involved as there are many more possible configuration transitions. One way to circumvent this, is to only consider totalistic rules, e.g.~when identifying global periodicity \cite{boccara1999totalistic}. Also recall the complexity trade-off between a CA with a larger neighbourhood and a CA with more states \cite{smith1971cellular}, which may aid to rephrase particular definitions. One such example is given in Fig.~\ref{fig:complexity-tradeoff2}.

When the neighbourhood is irregular (i.e.~for NAs) but the neighbourhood size is homogeneous, Li \cite{Li1992} showed that rule table parameter definitions can be generalised. This no longer holds when the local update rule does not operate on a neighbourhood with fixed size; this rule, then, is therefore almost always phrased in a totalistic fashion. Marr and Hutt \cite{Marr2005}, for example, determine a node state update based on the weighted density of states over all neighbouring nodes,
\begin{align*}
s(c_i,t+1) = \phi\left(\dfrac{1}{|\mathcal{N}(c_i)|}\sum\limits_{c_j \in \mathcal{N}(c_i)}A_{ij}s(c_j,t)\right).   
\end{align*}
This type of local update function allows for analysis, but not in the sense of the classical rule table. A similar consideration holds for global properties of the NA, and it is generally advised to make use of tools and properties devised for the analysis of complex networks, in particular in relationship to the node degree distribution \cite{mata2020complex}.

Memory CAs can have a rule table, as we illustrated in Fig.~\ref{fig:rca_rules}. If additionally the memory CA is reversible, only half of the rule table contains all required information for constructing rule table parameters. However, if the memory degree $M$ is large, such a rule table may become impractically large, impeding an elegant generalisation of the rule table parameters in Tab.~\ref{tab:list-of-math-tools-phenotypic}. When previous states are weighed, the resulting memory CA of course runs into the same issues as the CCAs discussed in Section \ref{sec:multi-state_CA}.

As for CAs with a larger neighbourhood size, generally the space of possible (subsequent) configurations in a memory CA increases in size, resulting in more involved global properties. The particular case of the reversible CA, however, strongly limits possible dynamics: every configuration has one and only one successor and predecessor. This generally strongly simplifies global mathematical description \cite{morita2012reversible}.

\subsubsection{The ENCA phenotype}

Subsequent 1-D configurations of ENCAs with a larger neighbourhood can still be described with local properties defined for ECAs. Due to more distant (but still local) information transfer, spatial correlations are generally achieved on shorter time scales. We also observe that simply the size of the neighbourhood may affect the global behavioural class \cite{Kayama2016}.

Phenotypical classification and quantification of ENCAs in higher dimensions and -- especially -- with irregular tessellations, requires some attention. The behavioural classes that are employed for 1-D CAs generally apply for CAs in two or more dimensions \cite{packard1985twodimensional}. However, Gerling \cite{gerling1990classification} observed drastically different behaviour for 3-D CAs. Generally, the much larger variety of dynamical outcomes in higher dimensions suggests that a richer classification (i.e.~more than four classes) may be informative \cite{stauffer1989classification}.

In terms of local properties of dynamical behaviour (cf.~Tab.~\ref{tab:list-of-math-tools-phenotypic}) of $n$-D CAs, either the metric should be extended into $n$ dimensions, or some measure for direction should be provided. As an example for the first case, consider some local structure theory where a particular property is averaged over an $n$-D block or sphere. For the second case, imagine the Lyapunov exponent related to the spread of a difference pattern in one specified direction, which could be averaged over all directions \cite{vispoel2023lyapunov}, or directing according to some perturbation front \cite{courbage2006lyapunov}. Baetens and De Baets \cite{baetens2013topology} use Lyapunov exponents for totalistic CAs with irregular tessellations, demonstrating that even something as drastic as a phase transition can be induced by solely altering the topology.

Network automata disregard all sense of `geographical' direction. Behaviour on networks can however still be classified in the classical sense, e.g.~according to the Li-Packard classes \cite{Li1992}. In particular, by adjusting a number of entropy definitions to the context of a complex network, Marr and Hütt \cite{Marr2005} showed that outer-totalistic NAs on various network types can evolve according to all four Wolfram behavioural classes. They also observed that the fraction of such NAs that exhibit complex behaviour shrinks for increasing neighbourhood size \cite{marr2009outer}. At the same time, the analysis of emergent NA dynamics can be used to classify the type of network itself when the adjacency matrix is unknown, e.g.~by running a NA based on the Game of Life \cite{miranda2016}.

Local properties can still be investigated when the `direction' required in the original definition from a particular node is generalised as some well-defined path through the network. The generalisation of i.a.~the Lyapunov exponent \cite{luque2000lyapunov}, then, allows for quantifying the robustness of an NA to topological variations. NAs exhibit topology-induced phase transitions \cite{baetens2013topology}, and topology influences an NA's tendency towards self-organisation \cite{gershenson2012guiding}.

CAs with memory are typically characterised with the canonical behavioural classes. Interestingly, adding memory to ECAs can also induce a class transition. Martínez et al.~\cite{martinez2015dynamics} show that almost half of all ECAs will completely transform their behaviour when information from up to ten previous time steps is taken into account. In particular, adding memory to CAs based on elementary rules can help `design' complex behaviour \cite{martinez2010how}. This is confirmed by Ninagawa et al.~\cite{ninagawa2014phase}.

\subsection{Applications of ENCAs}

\subsubsection{Computational advances}

1-D binary CAs with a larger neighbourhood are typically more suited than ECAs to solve computational problems such as density classification, even though no finite-radius perfect solution exists \cite{land1995perfect}. This is because they allow a wider range of rules, and because they represent some `intermediate' model between a purely local and global model. In the context of cryptography, randomness properties of a 1-D CA have also been investigated \cite{oliveira2010exhaustive} for larger radii. In particular, $r=2$ (`pentavalent') CAs were shown to be advantageous for a type of plain text cryptography called stream cipher design \cite{lakracarpenter2018}, while still remaining computationally feasible.

Most applications of ENCAs, however, require more than one dimension and/or special tessellations. The performance difference between the (2-D) Moore and Von Neumann neighbourhoods has been reported for image cryptography \cite{abudalhoum2012digital}. Adamatzky \cite{adamatzky2006glider} showed that hexagonal grids can support logical operations by generating gliders reminiscent of the Game of Life, and also Penrose tilings were shown to support some variation on the Game of Life \cite{owens2010investigations}.

Turning to NAs, many computation-related questions remain open, but some advances have clearly been made. Watts \cite{watts2003} examined the majority and synchronisation problems for small-world networks, showing that a local majority rule was typically sufficient for solving the problem, despite being inadequate for regular networks \cite{Das1994}. Darabos et al.~\cite{DARABOS2007} considered a wider range of network architectures, but still concluded that a small-world network is superior for solving the density or synchronisation problem.

Memory CAs, finally, also appear to facilitate density classification \cite{alonsosanz2013cellular}, although more research into their relation to computational problems is required.

\subsubsection{ENCAs in mathematical modelling}

Nearly all CA applications in mathematical modelling require a CA with more than one spatial dimension. Some authors have explicitly examined the influence of 2-D tessellations on model properties. Hopman and Leamy~\cite{hopman2010triangular}, for example, compared irregular triangular tessellations in elastodynamics with the rectangular case \cite{leamy2008application}. Also for geographical applications such as urban growth modelling, triangular \cite{ortigoza2015unstructured} and irregularly-shaped CAs \cite{stevens2007gis} are routinely applied. In social sciences, Flache showed that many general properties are robust to variation in the grid structure \cite{flache2001irregular}.

Mathematical modelling has a rich history involving complex networks, with networks optimised for representing i.a.~metabolic \cite{Jeong2000}, social \cite{Taylor2015}, and biological \cite{Doye2005} networks. When no direct access to the network's adjacency matrix is available, an analysis of the dynamical behaviour induced by an NA may serve as a tool for categorising the network type, useful for e.g.~authorship attribution of written text \cite{machicao2016}. With this aim, Miranda et al.~\cite{miranda2016} proposed an NA running an extended definition of the Game of Life, using measures such as Shannon entropy and word entropy in the decision process. This work was refined more recently by Ribas et al.~\cite{ribas2020lifelike} and Zielinski et al.~\cite{zielinski2022}. NAs are of course also used when, inversely, the network is known but the emergent dynamical behaviour is not. Such applications likewise cover a wide spectrum of research areas, ranging from urban planning \cite{osullivan2001graph} to fungal growth \cite{smith2011network}.

For memory CAs, finally, the most obvious application of memory CAs is again found in their relation to reversibility and (hence) conservation \cite{kari2018reversible}. This property justifies its use in modelling e.g.~Newtonian mechanics \cite{alonso2008cellular}, although also here the number of practical applications found in the literature is limited.

\section{Non-uniform CA}
\label{sec:non-uniform_CA}

\subsection{Motivation and definition}

A final extension to CAs allows for different cells to follow different rules, i.e.~the cells behave in a `non-uniform' fashion. CAs in which rule uniformity is abandoned are therefore called non-uniform CAs (or `heterogeneous' CAs, e.g.~\cite{gao2020urban}),  often abbreviated as $\nu$CAs. Now, the local update rule depends on the cell identity $i$ and time step $t$, so we have $\phi = \phi(i,t)$. Cattaneo et al.~\cite{cattaneo2009} identify three reasons to relax the uniformity constraint. First, generality, as some models may require local interactions to be dependent on some global spatial positioning. Second, the investigation of structural stability, because some emergent properties may be strongly dependent on all cells following the same rule, implying that some of the CA behaviour is `artificial'. Third, reliability (robustness to noise), particularly in the context of fast parallel computation. These motivations are highly similar to those for ACAs and SCAs, and we will in fact show that in many cases the mathematics is similar (or identical) as well.

In principle, $\nu$CAs may have an arbitrary number and type of local update rules that depend on time and space. The simplest non-trivial size-$N$ $\nu$CA allows for two local elementary update rules, say $\phi$ and $\psi$. We arrive at the following definition (and at the final leg in Fig.~\ref{fig:CA-family-diagram}):
\begin{definition}[Non-uniform cellular automaton]
\label{def:nuCA}
A non-uniform cellular automaton ($\nu$CA) is identical to an ECA (Def.~\ref{def:ECA}), with the exception of the sixth property:
\begin{align}
    s(c_i, t+1) = \begin{cases} \phi(\tilde{s}(\mathcal{N}(c_i), t)), & \text{if } i \in Z(t),\\
    \psi(\tilde{s}(\mathcal{N}(c_i), t)), & \text{else},
    \end{cases}
\end{align}
where the set $Z(t) \subseteq \{1, \ldots, N\}$ generally changes over time and determines the `rule allocation'.
\end{definition}
We refer to $\vert Z(t) \vert / N = \nu(t) \in [0,1]$ as the degree of non-uniformity, where $\nu = 1/2$ represents maximal non-uniformity. Note that if $\psi$ is the identity rule ($W(\psi)=204$), this is equivalent to the general definition of $\alpha$-asynchronous CAs (Def.~\ref{def:ACA} and Section \ref{sec:asynchronous_CA}), with $\nu = \alpha$. In that spirit, the set $Z(t)$ again allows for distinguishing between various types of $\nu$CAs.

\subsection{Non-uniform CA taxonomy}

\subsubsection{Spatial $\nu$CA variations}

If $Z(t) = Z$, the rule allocation is static. Clearly, for a two-rule size-$N$ $\nu$CA, there are $2^N$ possible rule allocations in total, and $\dbinom{N}{\vert Z \vert}$ for a particular $\nu$. We may, however, distinguish between various types. 

First, a stochastic rule allocation, for which every cell has a probability $p$ (resp.~$1-p$) of following rule $\phi$ (resp.~$\psi$), such that $\vert Z \vert \sim \text{Bin}(N,p)$, and $\text{E}[\nu] = p$. This of course is identical to the general definition of an SCA (Def.~\ref{def:SCA}). This setup is of interest because, especially for $p=1/2$, it is the `least global' non-uniform rule allocation, which is still in line with the CA philosophy of maximal locality \cite{bhattacharjee2020survey}, and which generally still allows for mathematical generalisations. As with SCAs, we may additionally demand that $|Z(t)|$ is constant.

Second, only a single cell may behave differently, i.e.~$Z = \{j | 0 < j \leq N\}$. This is of interest in the research of robustness to noise due to permanent failures \cite{dennunzio2014three}, allowing the quantification of the effect of one `rogue cell'.

Third, $Z$ can be chosen such that the rule allocation is periodic, e.g.~$Z = \{2j\ \vert\ 0 < j \leq N/2\ \land j \in \mathbb{N}\}$, or more generally
\begin{align}
    Z = \bigcup\limits_{k=0}^{N/T-1}\{1+kT, \ldots, \nu T+kT\}.
\end{align}
Here $T$ represents the period, in such a way that $N$ is a whole multiple of $T$, and $\nu = n/N$ with $n \in \{0, \ldots, N\}$. This approach represents a setup that more conveniently enables properties with a closed mathematical expression.

The fourth type is the most general one, where $Z$ is entirely `user-defined' at the global level. One notable case that is especially relevant for mathematical modelling may be called \textit{clustered} rule allocation, for which adjacent cells are on average more likely to follow the same rule. Such clusters may be generated stochastically, but typically they are defined by the modeller based on a spatial distribution of e.g.~geographical features. For instance, in a wildfire model \cite{Alexandridis2011}, this may be understood as geographical regions containing water versus those containing soil, for which different rules apply.

The CA's non-uniformity can be further increased, going beyond Def.~\ref{def:nuCA}. This is done by including more rules (that need not be elementary), which generally implies more variation and less mathematical oversight. Three mathematical types of these `general' $\nu$CAs allow for the generalisation of properties of uniform CAs, so-called default-rule, structural-period, and radius-$r$ $\nu$CAs \cite{dennunzio2012nonuniform}. As all three definitions require some symmetry around a central cell, the artificiality of their setup generally impedes any interesting application beyond mere mathematical constructions.

\subsubsection{Temporal $\nu$CA variations}

If $Z(t)$ is time-dependent, we may distinguish between purely temporal and spatiotemporal non-uniformity.

\paragraph{Purely temporal $\nu$CAs}

If (for binary $\nu$CAs) $Z(t)$ identifies at time $t$ either all cells or none of the cells, the CA is spatially uniform, but its global update function can change over time. For a two-rule $\nu$CA, this may be expressed as a set $\widetilde{Z} \subseteq \{0, \ldots, t_\text{f}\}$ with $2^{t_\text{f}+1}$ possible choices, where $t_\text{f}$ is the final time step in the simulation. We may identify the same general variations as for the spatially non-uniform case, with a similar motivation. Note that the stochastic case, where $\widetilde{Z} \sim \text{Bin}(t_\text{f}+1,q)$ determines whether global rule $\Phi$ (reps.~$\Psi$) applies with probability $q$ (resp.~$1-q$) is entirely equivalent to a temporal SCA \cite{roy2022temporally} (cf.~Section \ref{subsubsec:temporal-SCA-variations}).

\paragraph{Spatiotemporal $\nu$CAs}

In the most general case, the rule allocation changes both in space and in time. We discuss two special cases of two-rule $\nu$CAs with some systematic local rule allocation, and demonstrate how this locality causes the $\nu$CA to be equivalent to a uniform (multi-state) CA, and vice versa.

First, $Z(t)$ may change over time if the local rule depends on the neighbourhood. However, as an example, suppose rule 153 applies when the neighbourhood is symmetric, and rule 102 otherwise. In binary code, we write this as $153 = \mathbf{1}0\mathbf{0}11\mathbf{0}0\mathbf{1}_2$ and $102 = 0\mathbf{1}1\mathbf{00}1\mathbf{1}0_2$, where the bold bits indicate when the rule applies. The effective uniform rule is then simply the combination of the boldface bits, so $11000011_2 = 195$. This is easily generalised to CAs with more states and/or a larger neighbourhood.

Second, we can build a simple $\nu$CA with local rule allocation dynamics, stacking from two rules $\phi$ and $\psi$. Suppose rule $\phi$ applies to cells $\{c_i | i \in Z(t)\}$, and rule $\psi$ applies to all others. From some initial rule allocation $Z(0)$, the set $Z(t)$ is updated according to some local rule, say another Wolfram rule $\xi$ with
\begin{align}
    [\phi_i(t+1) = \phi] = \xi([\phi_{i-1}(t) = \phi], [\phi_i(t) = \phi], [\phi_{i+1}(t) = \phi]).
\end{align}
Here $\phi_i(t)$ indicates the rule that updates cell $c_i$ at time step $t$, and $[\phi_i(t) = \phi]$ outputs $1$ (resp.~$0$) when the update rule is rule $\phi$ (resp. rule $\psi$).

Note, now, that any cell can be in one of four `dual' states, namely $\{0, \phi\}, \{0, \psi\}, \{1, \phi\}$ or $\{1, \psi\}$, and the dual state of a cell at the next time step is deterministically and uniformly dictated by the dual states of its neighbourhood. The dual state's first element depends on the first elements of the dual states of the cell's neighbourhood, as well as the second element of the dual state of the cell itself. The dual state's second element depends only on the second elements of the dual states of the neighbourhood. This amounts to $256^3 \sim 10^7$ possible rules, essentially picking a Wolfram rule for $\phi, \psi$, and $\xi$. These rules form a (tiny) subset of the $\sim 10^{38}$ quaternary $\vert \mathcal{N}\vert=3$ local update rules. That is to say: a binary three-neighbourhood two-rule $\nu$CA can be mapped to -- and therefore simulated by -- a quaternary three-neighbourhood uniform CA. The general case of this mapping was demonstrated by Kamilya et al.~\cite{kamilya2021simulation}.

\begin{figure}
    \centering
    \includegraphics[width=.8\linewidth]{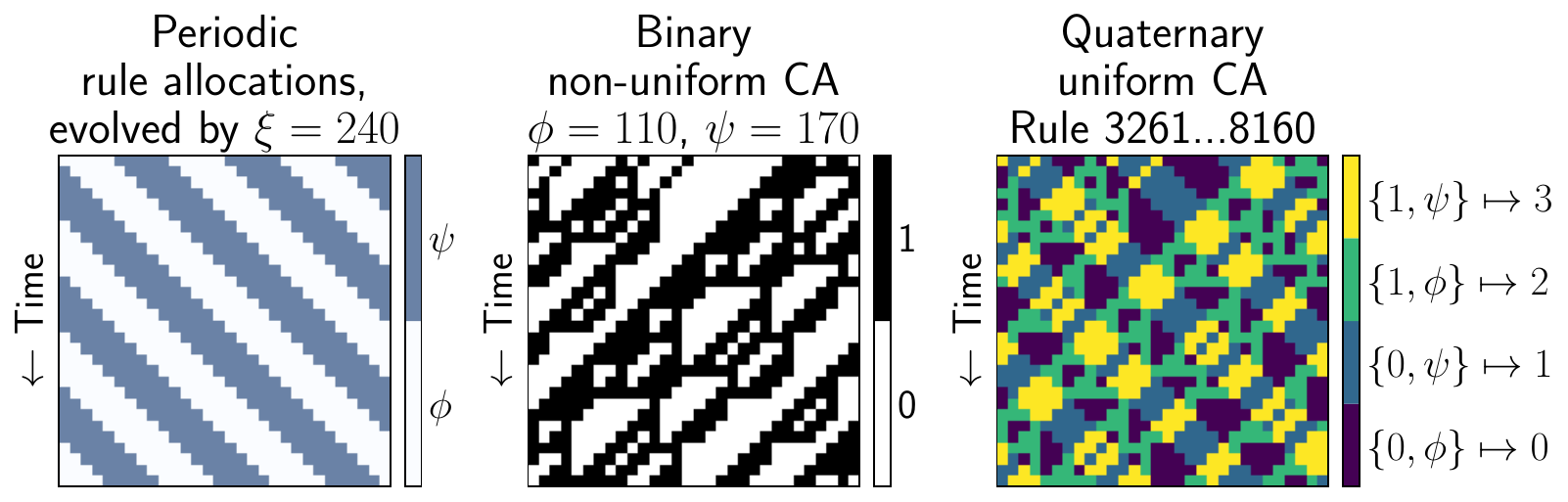}
    \caption{A four-state (quaternary) CA can simulate a $\nu$CA with locally defined dynamic rule allocation. Here, the cells in a $\nu$CA are evolved according to either rule $\phi=110$ or rule $\psi=170$. The allocation of these rules to the cells at a particular step is determined by the initial rule allocation -- in this case periodic with period $T=5$ --, and evolved by rule $\xi=240$ (left). The resulting $\nu$CA spacetime diagram is shown in the centre diagram. The righthand spacetime diagram displays the quaternary uniform CA that simulates this $\nu$CA by means of rule \num{326159115819648357517613529732932608160}.} 
    \label{fig:nuCA-simulated-by-quat-CA_phi110-psi170-xi240}
\end{figure}

As an example of such a mapping, consider a binary two-rule $\nu$CA $\mathcal{C}_1$ and a quaternary uniform CA $\mathcal{C}_2$, both with $\vert\mathcal{N}\vert = 3$. If we consider $\mathcal{C}_1$ to have four dual states, we may perform a mapping (choosing one of $4!$ possibilities) between the states of both CAs:
\begin{align*}
    (\{0, \phi\}, \{0, \psi\}, \{1, \phi\}, \{1, \psi\}) \mapsto (0, 1, 2, 3).
\end{align*}
This implies that a cell in an even state in $\mathcal{C}_2$ corresponds to a cell that obeys rule $\phi$ in $\mathcal{C}_1$. A cell in a state smaller than 2 in $\mathcal{C}_2$ corresponds to a cell in state 0 in $\mathcal{C}_1$. These are two independent binary observations (odd/even and smaller/greater than 2), which always lead to a definite state value for the cell in the quaternary CA. We show an example of the result from this mapping in Fig.~\ref{fig:nuCA-simulated-by-quat-CA_phi110-psi170-xi240}.

More generally, a $\nu$CA governed by $l$ rules and hosting $k$ states can be mapped to a uniform $(k \times l)$-nary CA with the same neighbourhood size. The inverse is generally not true as the rule space cardinality of the former is much smaller than that of the latter:
\begin{equation*}
    \forall k, l > 1: k^{lk^{\vert \mathcal{N}\vert}} l^{l^{\vert \mathcal{N}\vert}} \ll (kl)^{(kl)^{\vert \mathcal{N}\vert}}.
\end{equation*}
This is due to the fact that many quaternary rules are illegal for $\nu$CA simulation, because the outputs of all $(kl)^{\vert \mathcal{N}\vert}$ neighbourhood configurations must be mutually consistent.

\paragraph{Hybrid CAs}

Because one of the central CA properties is its locality, CAs that violate this attribute are sometimes called `hybrid CAs'. The local update rule can generally be written as $\phi = \phi(\tilde{s}(\mathcal{N}, t), \theta(i,t))$, with $\theta$ a time- and space-dependent global parameter. This essentially allows for an infinite number of rules and is arguably the most general form of a $\nu$CA. Its merit is however again mostly in applications, as closed mathematical expressions are now generally infeasible.

\subsection{Non-uniform CA genotype and phenotype}
\label{subsec:nuCA_properties}

\subsubsection{The non-uniform CA genotype}

Dennunzio et al.~\cite{dennunzio2014three} identify three directions in the genotypical characterisation of $\nu$CAs: (1) the investigation of sensitive dependence on initial conditions, (2) the generalisation of conditions for dynamical properties such as number conservation and reversibility, and (3) the exploration of the set of reachable states, and of the fixed-point sets in particular. Because many genotypical properties of classical CAs are supported by the uniform application of a single rule table, many of these are no longer true for $\nu$CAs in general. Some clever generalisations can be made for particular $\nu$CAs \cite{dennunzio2012nonuniform}, but these results are difficult to exploit outside the realm of pure mathematics.

Dependence on initial conditions, which is a measure of chaoticity, has been studied from a genotypical perspective by Kamilya and Das \cite{kamilya2019study} by quantifying the information flow through the CA.

Rule table parameters for $\nu$CA may be constructed from the rule table parameters associated with the rules that make up the $\nu$CA, e.g.~by simply taking the average. However, an approach that depends less on contingent choices is supported by the observation that the uniform $k$-nary CA that simulates the $\nu$CA at hand, must also encode its genotype. That is to say: just look at the rule table of the uniform CA that simulates the $\nu$CA. Whether or not this approach is helpful, however, depends on whether (1) this complexity trade-off map between the two CA families exists -- which is not true in general --, and whether (2) the genotype of the $k$-nary CA is known and sufficiently informative. If these conditions are met, a $\nu$CA can be associated with all rule table parameters and global properties known to the $k$-nary CA, independent of the initial configuration. To our knowledge, no such approach has yet been taken.

Regarding the identification of genotypical dynamical $\nu$CA properties (Dennunzio et al.'s second research direction), much theoretical work has been focused on number conservation (e.g.~\cite{salo2014realization}). In particular, Wolnik et al.~\cite{Wolnik2023} recently identified all number-conserving and reversible finite 1-D $\nu$CAs constructed from Wolfram rules. More generally, Dennunzio et al.~\cite{dennunzio2013local} formally investigated properties of $\nu$CAs for different rule allocations, identifying a number of complexity classes, but also again pinpointing the many methodological challenges that non-uniformity invokes \cite{dennunzio2013local}.

Concerning the third research direction, Dennunzio et al.\cite{dennunzio2014three} have applied de Bruijn graphs in formal language theory to identify a number of fixed-point configurations of the $\nu$CA. Adak et al.~\cite{adak2017synthesis} developed a construction method for $\nu$CAs whose fixed-point set has cardinality one, i.e.~that have a static end state. Taking a slightly different perspective, Adak et al.~\cite{adak2021reachability} recently developed an (often efficient) algorithm that decides whether or not a $\nu$CA configuration can be reached from another one.

\subsubsection{The non-uniform CA phenotype}

Little work has been done on the classification of $\nu$CAs based on their phenotype. Arguably, this is due to the large number of possibilities -- allocating two rules in a periodic CA of $N$ cells provides $2^N/N$ options --, and due to the unpredictable interference of rules -- e.g.~two periodic rules may cooperate, generating a complex pattern. While the latter reason is indeed a symptom of the theoretical difficulty associated with $\nu$CAs, it is also an important reason to investigate this model further. After all, it demonstrates yet again the principle of emergence in (evolving) artificial life \cite{sipper1994evolution}.

We argue that computational resources can now accommodate large-scale simulations that may uncover instructive statistical correlations within the large set of $\nu$CA possibilities. This is especially the case considering that almost all local phenotypical properties remain well-defined. Nonetheless, the amount of research in this direction is limited, and established results are mostly formulated in terms of equivalent stochastic CAs. In particular, the aforementioned work by Fat\`es \cite{Fates2017} on diploid CAs in fact also documents phase transitions in $\nu$CAs with random dynamic rule allocations, and Roy et al.'s work on temporally stochastic CAs \cite{roy2022temporally} demonstrates the emergence of unexpected behaviour in purely temporal $\nu$CAs.

As a closing remark, note that also the Lyapunov exponent remains well-defined, but that the introduction of non-uniformity opens the possibility of inspecting a new type of sensitivity: dependence on initial rule allocation. The initial defect is then represented by breaking uniformity via the allocation of a different rule to one cell, possibly causing a defect cone in the resulting spacetime diagram.

\subsection{Applications of non-uniform CAs}

\subsubsection{Computational advances}

In the context of computer science, $\nu$CAs were explored in the context of tasks such as the majority problem \cite{sipper1998computing}. In particular, Sipper \cite{sipper1996} showed that $\nu$CAs are capable of solving this problem in over 90 percent of the cases, mainly due to the higher number of design variations. As with the uniform case \cite{mitchell1994}, evolutionary algorithms can be of great help for identifying these rules, as well as for constructing a $\nu$CA that performs simple arithmetic operations \cite{grouchy2016evolving}. Further, Kumaravel and Meetei \cite{Kumaravel2013} used reversible $\nu$CAs for increasing the complexity (and hence the safety) of encryption-decryption algorithms. This research was continued more recently by Mukherjee et al.~\cite{mukherjee2021nonuniform}. A better understanding of the theoretical foundation is of particular interest for these approaches, and a lot is left to uncover.

\subsubsection{$\nu$CAs in mathematical modelling}

Outside theoretical computer science, $\nu$CAs have been applied to model both natural and man-made phenomena, such as epidemiology \cite{eosina2022covid}, insect movement \cite{Romano2006}, and biochemistry \cite{elsayed2017}. In theoretical biology, Adams \cite{Adams2022} again highlights the importance of randomness in evolution, by means of a purely temporal $\nu$CA where the local rule is determined by some global property of the system. Hybrid CAs have been used in wildfire modelling where the global parameter could be the time-dependent direction of the wind \cite{Alexandridis2011}, and in cancer therapy research where the changing nutrients concentration affects the local transition rule \cite{Ribba2004}. Another example of a hybrid CA is Vichniac's Ising model \cite{vichniac1984}, in which the temperature plays the role of global parameter that affects transition probabilities.

\section{Conclusion and outlook}
\label{sec:conclusion}

We reviewed literature on five CA families: asynchronous CAs, stochastic CAs, multi-state CAs, extended-neighbourhood CAs, and non-uniform CAs. Each of these families is characterised by a distinct departure from the definition of ECAs, and are interconnected by mathematical identities, as shown in this taxonomy's central diagram (Fig.~\ref{fig:CA-family-diagram}). Despite the fact that some CA types and applications were not covered, we consider this review to be highly comprehensive. This is because we chose to only leave out particular topics when the relation to ECAs was considered too remote (e.g.~spatially continuous CAs), when the CA was too idiosyncratic to a particular research field (e.g.~quantum CAs), or when applications were not chiefly of a scientific nature (e.g.~CAs in the arts).

This survey disclosed a number of noteworthy patterns and realisations within the study of CAs. First, we emphasise the conceptual paradox that many CA families and types are in some sense opposed to the CA philosophy of strict locality, as often some global or random choice is imposed onto the individual cells. Second, we observe a trade-off of sorts: increased realism and applicability almost invariably comes at the cost of decreased mathematical elegance and methodological generality. Third, the methodological approach to each CA family brought with it an appreciation for the substantial mathematical overlap between these families. This clearly allows for an entire spectrum of approaches to a particular mathematical matter, but also bridges the gap between various academic branches within this field of research.

Concluding, we perceive the study of CAs as a great scientific opportunity, both as the object of mathematical inquiry as well as a modelling tool. CA research has a mature theoretical basis, presents many promising research avenues, and is studied at a time when computational power allows for thorough exploration. Non-uniform CAs and network automata in particular are currently well-positioned for disclosing their properties and merits. This is underlined by their close relationship to well-documented CA families and their properties, the many straightforward simulations that can be statistically described with these properties, and the wide spectrum of applications.

\section*{CRediT authorship contribution statement}

\textbf{Michiel Rollier:} Conceptualization, Methodology, Software, Writing - Original Draft, Writing - Review \& Editing. \textbf{Kallil M.C.~Zielinski:} Writing - Original Draft. \textbf{Aisling J.~Daly:} Conceptualization, Supervision, Writing - Review \& Editing. \textbf{Odemir M.~Bruno:} Supervision. \textbf{Jan M.~Baetens:} Conceptualization, Funding acquisition, Supervision, Writing - Review \& Editing.

\section*{Declaration of competing interest}

The authors declare that we have no known competing financial interests or personal relationships that could have appeared to influence the work reported in this paper.

\section*{Data availability}

No data was used for the research described in the article.

\section*{Acknowledgements}

This work has been partially supported by the FWO grant with project title ``An analysis of network automata as models for biological and natural processes'' [3G0G0122], and by the FWO travel grant with file name V418024N.





\bibliographystyle{elsarticle-num-names} 

\begin{flushleft}
\setstretch{1.0}
\footnotesize
\bibliography{ca-taxonomy-review_elsevier-formatted}
\end{flushleft}






\end{document}